\documentclass[aps,prb, preprint]{revtex4-2}

\pdfoutput=1
\usepackage{graphicx}
\usepackage{float}
\usepackage[leftcaption]{sidecap}
\usepackage[fleqn,tbtags]{amsmath}
\usepackage{mathtools}
\usepackage{amsmath}
\usepackage{amssymb}
\usepackage{titlecaps}

\begin{document}


\title{Magnetic Order and Magnetic Excitations in FeTe: How Good is a Short-Range Heisenberg Model?}


\author{Joe Marshall}
\email[Corresponding Author:]{ joseph.marshall@durham.ac.uk}
\affiliation{Department of Mathematical Sciences, Durham University, Stockton Road, DH1 3LE, Durham, United Kingdon \looseness=-1}
\author{Chris A. Hooley}
\email{hooley@pks.mpg.de}
\affiliation{Max Planck Institute for the Physics of Complex Systems, Nöthnitzer Straße 38, 01187 Dresden, Germany \looseness=-1}


\date{\today}

\begin{abstract}
We revisit the fitting of the Fang-Bernevig-Hu model [\textit{Europhys.\ Lett.\ }\textbf{86}, 67005 (2009)] to inelastic neutron scattering data carried out by Lipscombe \textit{et al}.\ [\textit{Phys.\  Rev.\  Lett.\ }\textbf{106}, 057004 (2011)].  We demonstrate that there are many quite different parameter choices within the experimentally observed phase (AFM3) that provide approximately equally good fits to the neutron data.  We note that (a) all of these parameter sets lie very close to a point of transition between the AFM3 phase and one of its neighbors in the classical phase diagram, and (b) many of them involve rather large values of the third-neighbor coupling $J_3$.  In light of these observations, we discuss whether the modeling of FeTe may need revision to allow for a slightly non-AFM3 ground state, orbital as well as spin physics on the Fe sites, the effects of electron itinerancy, or a combination of these.
\end{abstract}


\maketitle
\newpage

\section{Introduction \label{sec:intro}}
Iron telluride became a material of interest following the discovery of high-temperature iron-based superconductivity in 2006 \cite{kamihara2006iron, kamihara2008iron,johnston2010puzzle}. While FeTe is not itself superconducting, doping with sulfur or selenium ($\text{FeS}_{x}\text{Te}_{1-x}$ with $0.1\leqslant x \leqslant 0.5$ \cite{ma2022superconductivity} and $\text{FeSe}_{x}\text{Te}_{1-x}$ with $0.05\leqslant x \leqslant 1$ \cite{MARTINELLI2016phase, kreisel2020remarkable} respectively) leads to a superconducting state. The magnetism of the undoped compound, in particular the manner in which long range order is altered then destroyed as superconductivity emerges \cite{zhang2009density}, is thought to be important for understanding the mechanism of superconductivity. Spin waves in FeTe have been studied in the hope that they may provide insight into the origin of magnetic and superconducting effects in related materials \cite{lipscombe2011spin,fang2009theory}. It should be noted that FeTe is always synthesized with excess iron \cite{li2009first}. Throughout this paper we refer to ``FeTe'' for convenience even though, to be precise, one must have in mind Fe$_y$Te. In the sample subjected to inelastic neutron scattering \cite{lipscombe2011spin}, $y\approx1.05$.

Within this context, we calculate the spin wave dispersion relations predicted by the Fang-Bernevig-Hu model of FeTe \cite{fang2009theory}, and attempt to determine the coupling constants in that model by fitting our spin wave predictions to experimentally measured inelastic neutron scattering data. We use the data obtained by Lipscombe \textit{et al}.\  in 2011 \cite{lipscombe2011spin}; they also used the Fang-Bernevig-Hu model to make spin-wave fits to their data, but we believe that our approach --- based on a thorough classical analysis and a cost-function-based optimization of the fit --- is more systematic. 

The remainder of this paper is structured as follows. In section \ref{sec:Structure} we discuss the lattice structure of iron telluride and provide justification for the model adopted based on this structure. We also review what is known about the magnetic ground state of FeTe. Section \ref{sec:ClassicalPhase} reports on results obtained by studying the model in a classical limit, with emphasis on the phase diagram as a means of identifying regions of parameter space which correspond to the ground state. We highlight that the parameter space includes regions with smoothly changing incommensurate ordering wavevector near the observed ordering.

In section \ref{sec:SpinWaves} we move on to a spin wave analysis of the system, expanded about the observed ground state, and report our attempts to fit the Lipscombe \textit{et al}.\ data \cite{lipscombe2011spin}. We see in this section that there are a great many parameter combinations which provide equally suitable fits to the inelastic scattering data. We also find potentially unsatisfactory aspects in all of these fits which prompts us to ask whether the Heisenberg model used by Fang, Bernevig, and Hu may not describe the system as well as previously thought. Finally, we combine these results in section \ref{sec:Discuss} to highlight several ways in which the short-range Heisenberg model expanded about the observed magnetic ground state may fail to be suitable. Orbital effects, incommensurate ordering, and itinerant electron magnetism are suggested as possible sources of deviation from the predictions of the model used.

\section{Crystal Structure and Magnetic Model \label{sec:Structure}}

\begin{figure}[ht]
	\includegraphics[width=0.7\textwidth]{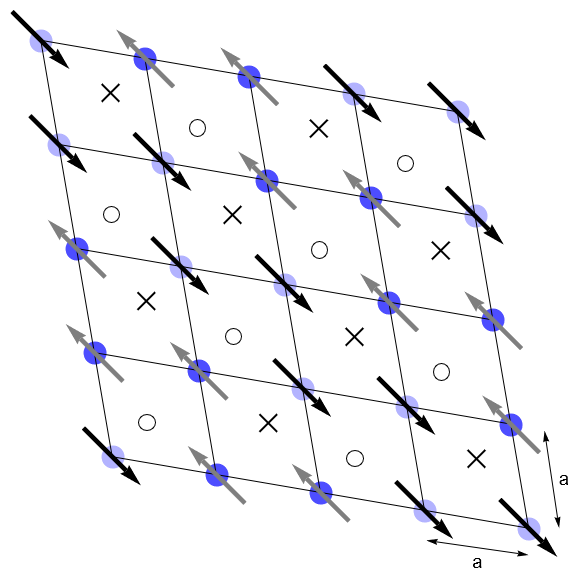}
	\caption{FeTe spin structure in the AFM3 classical ground state. Tellurium atoms mediate electron hopping between iron sites and are indicated by $\times$'s (above the plane) and $\circ$'s (below the plane). The distortion from a square lattice is accounted for by allowing for anisotropic nearest-neighbor and next-nearest-neighbor couplings. The lattice constant is $a$, which is the distance between neighboring iron sites.\label{fig:FullFeTe}}

\end{figure}

Iron telluride has a somewhat complicated lattice structure; however, only the iron sites are important for our analysis, with the tellurium atoms acting as a pathway for electron hopping \cite{turner2009kinetic}. Fig.\ \ref{fig:FullFeTe} depicts the planar structure of the lattice, with circles and crosses denoting tellurium atoms below and above the iron plane respectively. The static magnetic ordering, which will henceforth be referred to as the ``AFM3'' phase, is shown in this plane. Our analysis will focus on an individual plane of the crystal since coupling between iron layers is weak and the data we compare to has been integrated along the out-of-plane axis \cite{lipscombe2011spin}. We note that the lattice undergoes a distortion when magnetically ordered. This distortion is incorporated in the model by the allowance of anisotropic couplings. The coupling parameters are functions of the angle between iron sites \cite{fang2009theory} and the positions of tellurium atoms in the planes above and below. The model therefore treats the lattice as square, with effects of the distortion absorbed into the coupling constant values.

The spin wave dispersion is obtained by expanding the Hamiltonian about a classical equilibrium magnetic state, and is therefore only valid for parameters corresponding to that magnetic ground state. Hence, analysis of the model in the classical limit provides information on which combinations of coupling constants are feasible when fitting to the spin wave data. The magnetic ground state of the lattice can be observed via elastic neutron scattering and is, within error range, measured to have ordering wavevector $\mathbf Q=(\frac{\pi}{2a}$,$\frac{\pi}{2a})$ \cite{lipscombe2011spin} which we shall use to constrain parameter space. $a$ is the lattice constant for this system, defined in Fig.\ \ref{fig:FullFeTe}.

\begin{figure}[ht]
	\includegraphics[width=0.6\textwidth]{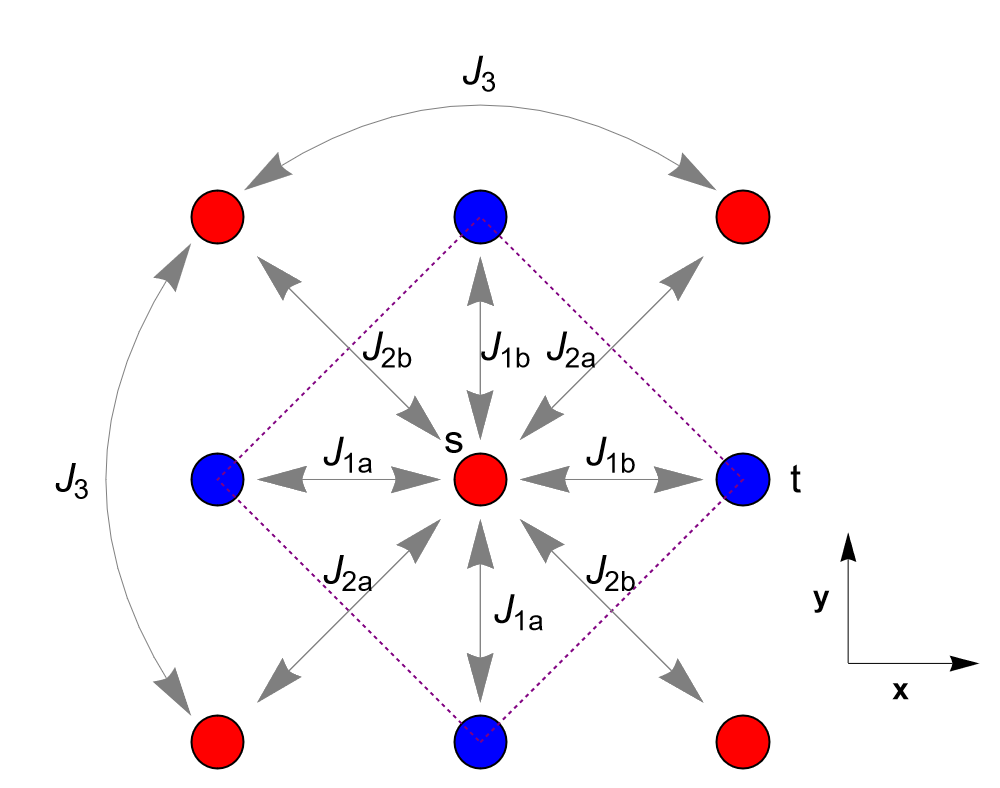}
	\caption{The structural unit cell with red $s$ sites and blue $t$ sites. The $s$ sites have $J_{1b}$ couplings in the positive $x$ and $y$ directions and $J_{1a}$ couplings in the negative directions. This is reversed for $t$ sites. There are five coupling constants in this model. $J_{1a}$ and $J_{1b}$ are nearest-neighbor couplings, $J_{2a}$ and $J_{2b}$ next-nearest-neighbor couplings, and $J_3$ the third-neighbor coupling. In this geometrically simplified model the origin of the disparity between the two nearest-neighbor couplings $J_{1a}$ and $J_{1b}$, and between the two next-nearest-neighbor couplings $J_{2a}$ and $J_{2b}$, is no longer obvious. We emphasize again that it arises from the distorted structure of the true crystal lattice. \label{fig:classcell}}
\end{figure}

Fig.\ \ref{fig:classcell} shows the unit cell of the Fang-Bernevig-Hu model for a square lattice with anisotropic nearest-neighbor and next-nearest-neighbor couplings, and isotropic third-neighbor couplings. The unit cell contains two sites, which we denote $s$ and $t$. We choose the site in the positive $x$-direction from the center to be contained in the cell. 
The Hamiltonian, in units such that spin operators are dimensionless and $J_{ij}$ have units of energy, is
\begin{equation}
H= \sum_{ i j } J_{i j} \mathbf S_i \cdot \mathbf S_j,
\label{eq:FBHHamiltonian}
\end{equation}
where the $J_{ij}$'s are as defined in Fig.\ \ref{fig:classcell} and zero for longer range couplings than those shown. In the classical limit, the operators become vectors, which we write as a unit vector multiplied by the spin magnitude: $\mathbf S_i \rightarrow \sqrt{S(S+1)} \; \mathbf s_i$. $\mathbf s_i$ will refer to the unit vector on an $s$ site and $\mathbf t_i$ that on a $t$ site. All $J_2$ and $J_3$ couplings are between like sites, while $J_1$ couplings link $s$- and $t$-type sites. Diagonalizing the Hamiltonian leads to:

\begin{equation}
H= S(S+1)\sum_{\mathbf p} \left[ \gamma_{\mathbf{p},s} |\tilde{\mathbf s}_{\mathbf p}|^2+\gamma_{\mathbf p,t} | \tilde{\mathbf t}_{\mathbf p}|^2\right],
\end{equation}
\begin{equation}
\gamma_{\mathbf p,s}=\gamma_{\mathbf{p},2}+\frac{1}{2} | \gamma_{\mathbf p,1}|,
\label{eq:gs}
\end{equation}
\begin{equation}
\gamma_{\mathbf p,t}=\gamma_{\mathbf p,2}-\frac{1}{2} | \gamma_{\mathbf p,1}|,
\label{eq:gt}
\end{equation}
\begin{equation}
\gamma_{\mathbf p,1}=J_{1a}(e^{i \mathbf p \cdot (\mathbf a+ \mathbf b)}+e^{i \mathbf p \cdot \mathbf a})+J_{1b}(e^{i \mathbf p \cdot \mathbf b}+1),
\end{equation}
\begin{equation}
\gamma_{\mathbf p,2}=J_{2a} \cos (\mathbf p \cdot \mathbf a)+J_{2b} \cos(\mathbf p \cdot \mathbf b)+2J_3 \cos(\mathbf p \cdot \mathbf a) \cos(\mathbf p \cdot \mathbf b).
\end{equation}
The Bravais lattice vectors in the above are $\mathbf a = a \mathbf x+a \mathbf y$ and $\mathbf b=a\mathbf x - a \mathbf y$ with lattice constant $a$. $\mathbf s_{\mathbf p}$ and $\mathbf t_{\mathbf p}$ are given by superpositions of the vectors which diagonalize the Hamiltonian:
\begin{equation}
\mathbf s_{\mathbf p}=\frac{1}{\sqrt{2}}(\tilde{\mathbf s}_{\mathbf p} + \tilde{\mathbf t}_{\mathbf p}),
\end{equation}
\begin{equation}
\mathbf t_{\mathbf p}=\frac{e^{i \theta_{\mathbf p}}}{\sqrt{2}}(\tilde{\mathbf s}_{\mathbf p}-\tilde{\mathbf t}_{ \mathbf p}),
\end{equation}
where $\theta_{\mathbf p}$ is the complex phase of $\gamma_{\mathbf p,1}$.
Finally, real-space unit vectors, $\mathbf s_i$ and $\mathbf t_i$, are the Fourier transforms of $\mathbf s_{\mathbf p}$ and $\mathbf t_{\mathbf p}$:
\begin{equation}
\mathbf s_i=\sqrt{\frac{2}{N}}\sum_{\mathbf p}e^{i \mathbf p \cdot \mathbf x_i} \mathbf s_{\mathbf p},
\end{equation}
\begin{equation}
\mathbf t_i=\sqrt{\frac{2}{N}}\sum_{\mathbf p}e^{i \mathbf p \cdot \mathbf x_i} \mathbf t_{\mathbf p},
\end{equation}
where $\mathbf x_i$ is the position of site $i$ and $N$ is the number of lattice sites. The sum is taken over the first Brillouin zone. 

The magnetic ground state for given parameters is determined by finding the global minima of the two branches of the structure factor $\gamma_{\mathbf p, (s,t)}$ and weighting the associated eigenvectors at the corresponding momentum/momenta in a way that satisfies the unit length constraints $\mathbf s_i \cdot \mathbf s_i=1$ and $\mathbf t_i\cdot \mathbf t_i=1$. We note that, from (\ref{eq:gs}) and (\ref{eq:gt}), $\gamma_{\mathbf p,t}\leqslant \gamma_{\mathbf p,s}$ for all $J$ and $\mathbf p$. Therefore, unless the two branches touch, one need only find the minima of $\gamma_{\mathbf p,t}$ and weight $\tilde{\mathbf t}_{\mathbf p}$ at the corresponding $\mathbf p$ value(s) subject to the above constraints. The two branches touch only when the conditions in one of the rows of Table \ref{tab:degenerateStructureFactor} are satisfied. 
\setlength{\tabcolsep}{3em}
\begin{table}[h]
\begin{tabular}{c|c}
$\mathbf p$ requirements    & $J$ requirements\\
\hline
$p_y=p_x \pm \frac{\pi}{a}$ & any values \\
$p_y=-p_x \pm \frac{\pi}{a}$  & $J_{1a}=J_{1b}$ \\
$p_y=-p_x$ & $ J_{1a}=-J_{1b}$
\end{tabular}
\caption{Conditions under which the two branches of the classical structure factor, $\gamma_{\mathbf p,s}$ and $\gamma_{\mathbf p, t}$, touch. \label{tab:degenerateStructureFactor}}
\end{table}

\section{Classical Phase Diagram of Iron Telluride \label{sec:ClassicalPhase}}
	\subsection{AFM3 Regions of Parameter Space \label{ssec:AMF3Regions}}

 We have already commented that the magnetic ordering wavevector can be determined entirely from $\gamma_{\mathbf p,t}$ except when $|\gamma_{\mathbf p,1}|=0$. In the previous section, we saw that this occurs on the entire Brillouin zone boundary when $J_{1a}=J_{1b}$ --- which is relevant given that the FeTe ordering vector is observed on the boundary. When the two branches touch at the global energy minimum, the Fourier-transformed on-site unit vectors $\mathbf s_{\mathbf p}$ and $\mathbf t_{\mathbf p}$ already diagonalize the Hamiltonian and as a result the orientation of spins on $s$ sites is independent from the orientation of spins on $t$ sites. Under these conditions, a minimum at the AFM3 point leads to a phase in which AFM3 is technically a ground state, but it is degenerate with any rotation of $t$ sites relative to $s$ sites. In reality, it is likely that the degeneracy is lifted at finite temperature or when $1/S$ corrections are included. 
 
\begin{figure}[h]
	\includegraphics[width=0.75\textwidth]{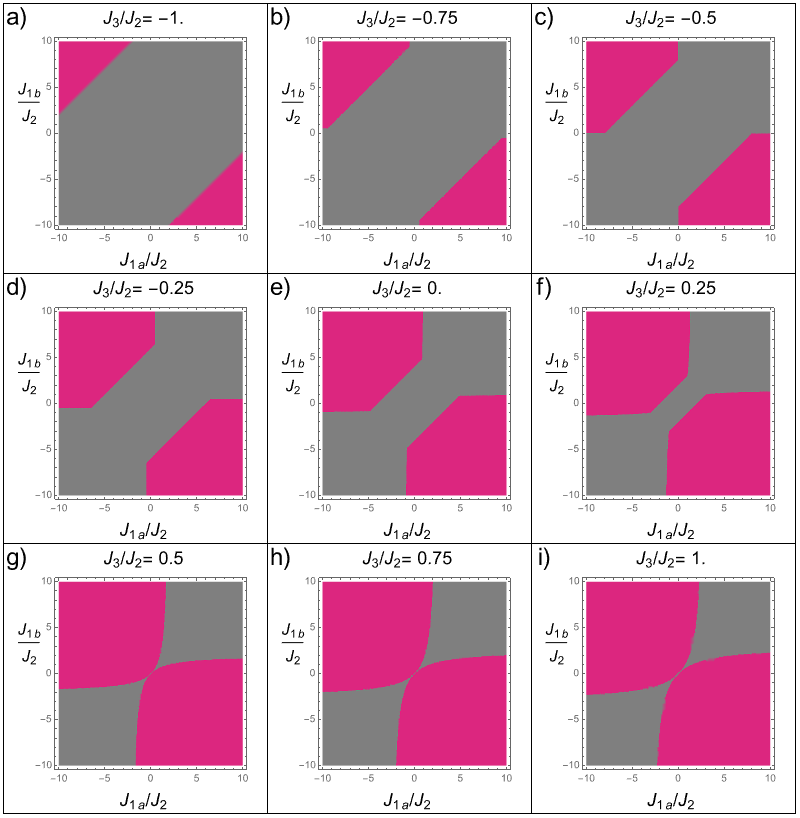}
	\caption{Classical phase diagram of the Fang-Bernevig-Hu model via plots of the $J_{1a}/|J_2|$ --- $J_{1b}/|J_2|$ parameter space. $J_{2a}=J_{2b}\equiv J_2>0$ for different $J_3/J_2$ values which label each plot. Pink indicates a region of parameter space in which the ordering wavevector is at ($\frac{\pi}{2a}$,$\frac{\pi}{2a}$), which is the experimentally observed AFM3 value. Non-AFM3 regions are shown in gray. \label{fig:classPhaseSpace}}
\end{figure}

We now proceed to determine parameter combinations for which the minimum is at  ($\frac{\pi}{2a}$,$\frac{\pi}{2a}$). This is not analytically viable; however, density plots of the parameter space, shown in Fig.\ \ref{fig:classPhaseSpace}, give a good idea of AFM3 accessibility when we fix $J_{2a}=J_{2b}>0$ (see section \ref{app:minusJ2} of the Supplemental Material for plots with anisotropic $J_{2a}$ and $J_{2b}$).
Pink regions indicate a structure factor minimum at the ordering wavevector corresponding to AFM3; however, note that a region would still be pink even if there were other degenerate minima. This can happen, for example, when $|J_{1a}-J_{1b}|=2(J_{2b}-J_{2a})$: in this case the structure factor has degenerate minima at $(\frac{\pi}{2a},-\frac{\pi}{2a})$ and $(\frac{\pi}{2a},\frac{\pi}{2a})$. However, such regions are measure zero in the parameter space and thus the plots are a good guide. 

Broadly, we observe that the AFM3 state primarily occupies regions in which $J_{1a}$ and $J_{1b}$ have opposite signs. This is unsurprising when considering the magnetic structure (i.e.\ in a double-stripe structure one of $J_{1a}$ and $J_{1b}$ must link aligned spins and the other must link anti-aligned spins); however, without orbital considerations it is hard to explain why this should be the case purely from the lattice structure. Additionally we notice that $J_3$ has a significant effect, with a more positive $J_3$ resulting in a larger proportion of the parameter space plane corresponding to the AFM3 state. $J_3$ couplings require twice as many electron hops between sites as $J_1$ and $J_2$ couplings and therefore in an insulating model we would expect $J_3$ couplings to be small in comparison.

We will see in section \ref{sec:SpinWaves} that fitting spin wave dispersion curves to the data does suggest a slight preference for large $J_3$. We note that for $J_{1a}$ and $J_{1b}$ to have the same sign while still giving AFM3 order $J_3$ should be more positive. Even then, the allowed values are fairly anisotropic, e.g. $J_{1a}/J_2 >4$  and $J_{1b}/J_2<1$ for $J_3=0$. If $J_3$ is large, then $J_{1a}$ and $J_{1b}$ can be approximately equal provided they are small. In summary there is a trade-off between isotropic $J_{1a}$ and $J_{1b}$, and small $J_3$. 

	\subsection{Near-AFM3 Phases}
The minimum of the structure factor, based on experimental observations via elastic neutron scattering \cite{lipscombe2011spin}, is at $(\frac{\pi}{2a} , \frac{\pi}{2a})$ in FeTe. Typically, the error in a measurement of this kind will be in the range 0.1\% - 1\% \cite{Huxley2023}. It is feasible that the true magnetic ordering wavevector could be off $(\frac{\pi}{2a},\frac{\pi}{2a})$ so long as it is within the error range of the measurements, if such states are theoretically accessible. By studying the structure factor we notice that it is indeed possible for a minimum to lie anywhere on the lines $(\frac{\pi}{a}-\epsilon,\epsilon)$ and $(\epsilon,\epsilon)$ with $0<\epsilon<\frac{\pi}{2a}$, meaning there are accessible states with ordering arbitrarily close to AFM3. 

\begin{figure}[ph]
	\includegraphics[width=0.5\textwidth]{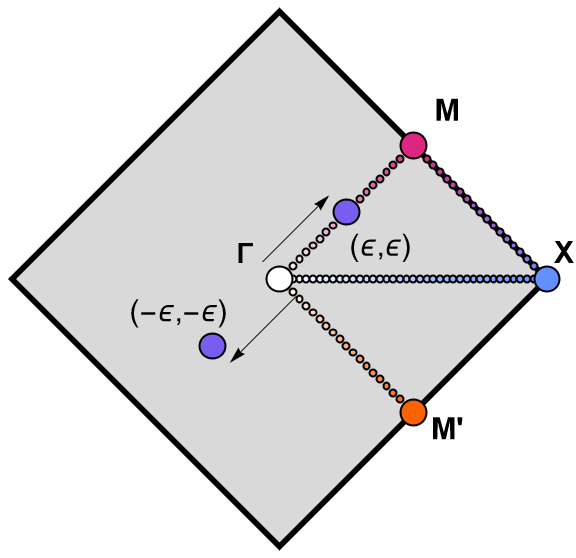}
	\caption{The structural Brillouin zone with the zone center ($\Gamma$), upper center-edge ($M$ --- AFM3), corner ($X$), and lower center-edge ($M'$) indicated. \label{fig:classBZCenter} }
\end{figure}

\begin{figure}[ph]
	\includegraphics[width=\textwidth]{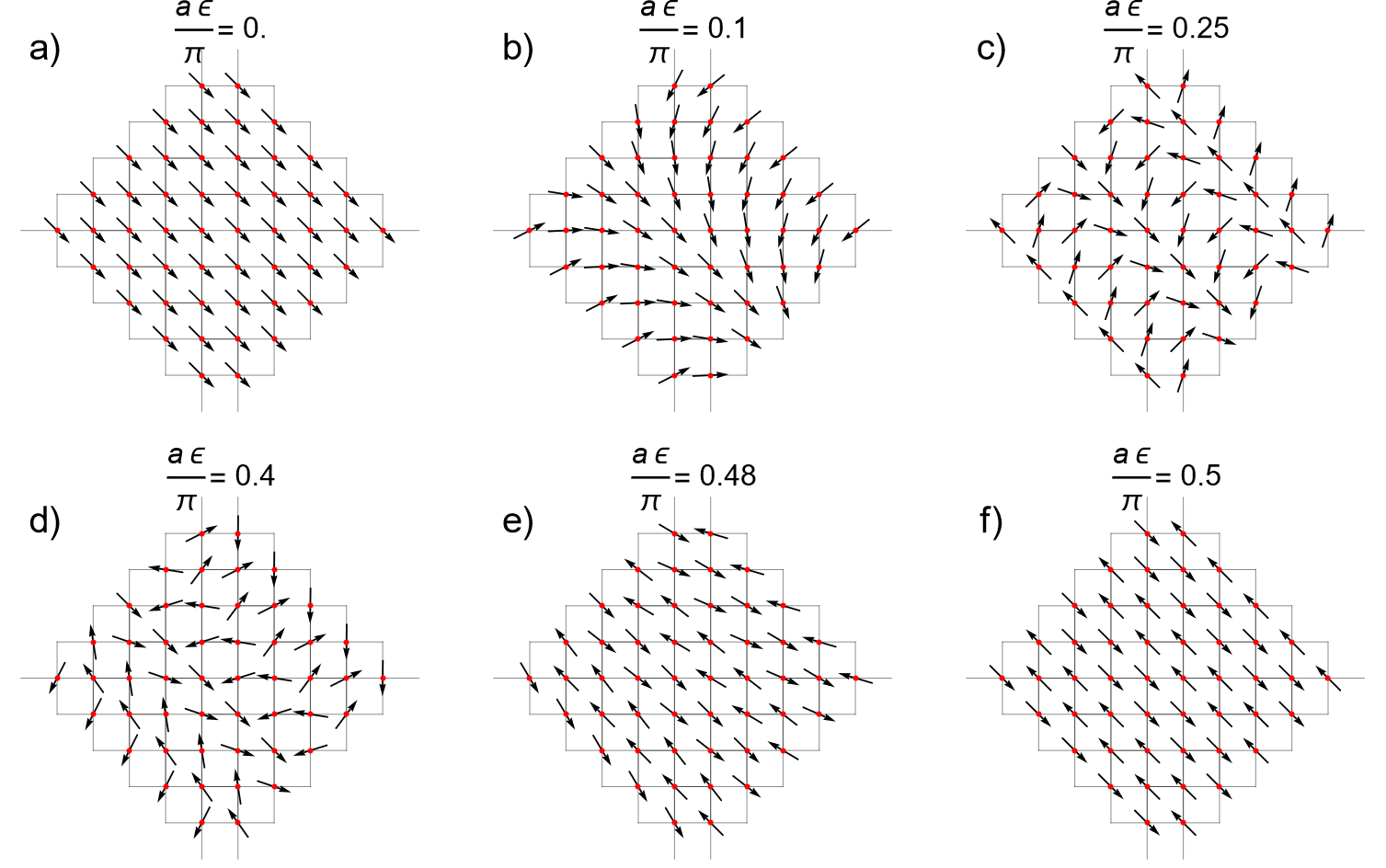}
	\caption{Real-space magnetic ordering resulting from a structure factor minimum at $(\epsilon,\epsilon)$. In this plot $J_{1a}=2J_{1b}=1$. This results in a ferromagnet for $\epsilon=0$ and the AFM3 state for $\epsilon=\pi/2a$. \label{fig:nearAFM3OtCP} }
\end{figure}

\begin{figure}[h]
	\includegraphics[width=\textwidth]{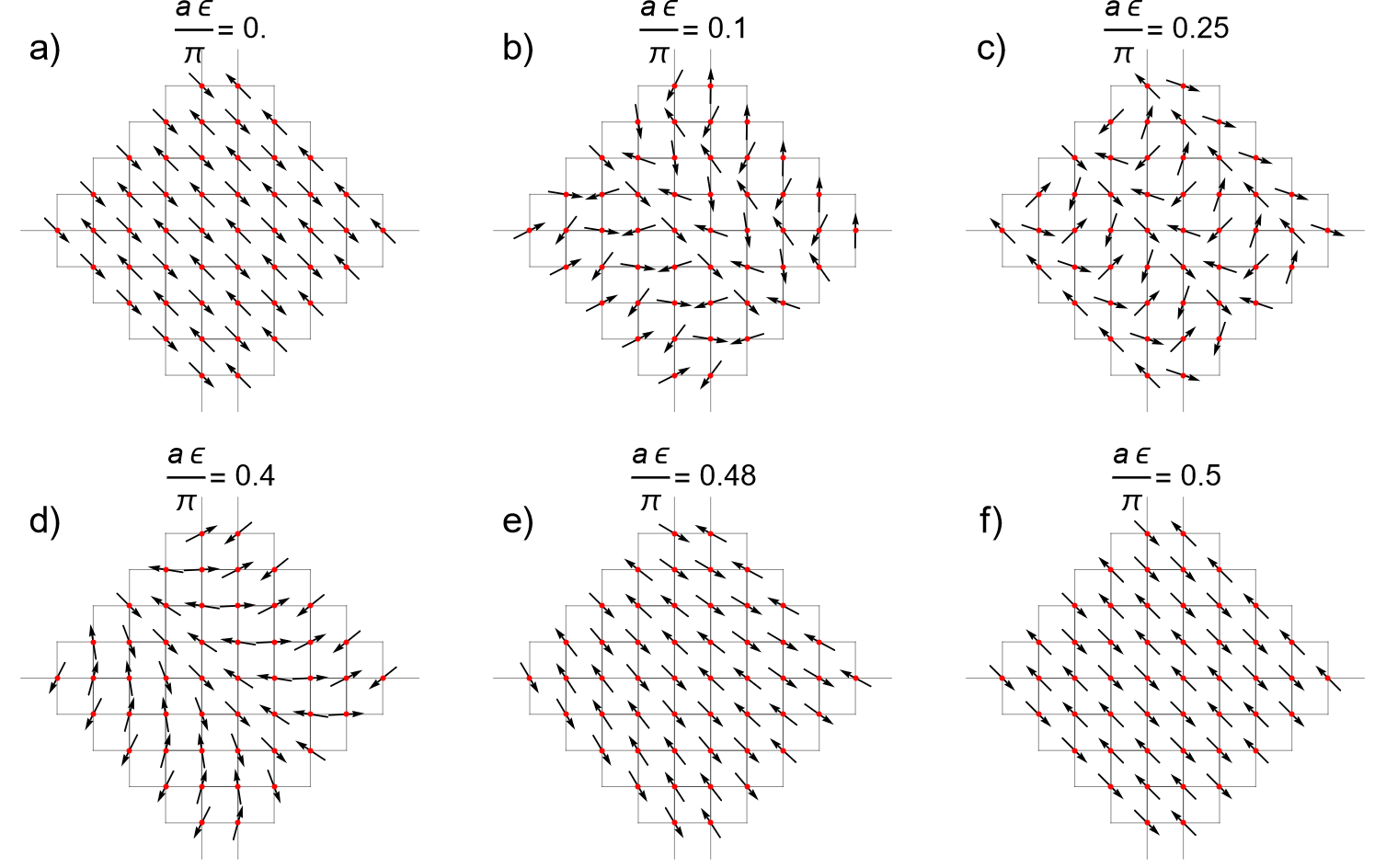}
	\caption{Real-space magnetic ordering as in Fig.\ \ref{fig:nearAFM3OtCP} except with $J_{1a}= - \frac{1}{2}J_{1b}=1$. This results in an anti-ferromagnet for $\epsilon=0$ and the AFM3 state for $\epsilon=\pi/2a$.\label{fig:nearAFM3OtCM} }
\end{figure}

Due to the reflection symmetries of the structure factor about $p_x=p_y$ and $p_x=-p_y$ any minimum  implies other minima at its reflection points, which must be equally weighted in determining the classical order. Fig.\ \ref{fig:classBZCenter} shows the Brillouin zone with these points marked for the $(\epsilon,\epsilon)$ case. Note that along this line the magnetic ordering actually depends explicitly on the relative values of $J_{1a}$ and $J_{1b}$. This had to be the case as we know a structure factor minimum at the origin ($\epsilon=0$) could indicate either a uniform or N\'{e}el state depending on the signs of $J_{1a}$ and $J_{1b}$. Fig.\ \ref{fig:nearAFM3OtCP} shows the evolution of the real-space magnetic ordering pattern along the line in the Brillouin zone for $J_{1a}$ and $J_{1b}$ such that the pattern at the origin is uniform, whereas Fig.\ \ref{fig:nearAFM3OtCM} shows the same evolution starting from a N\'{e}el state. Of course, $J_{1a}$ and $J_{1b}$ cannot be freely varied; one must choose them (along with $J_{2a}$, $J_{2b}$, and $J_3$) such that the global structure factor minimum remains at the same value of $\epsilon$ which greatly constrains possibilities.


\begin{figure}[ht]
\hspace*{1cm}
\vspace*{-.2cm}
	\includegraphics[width=0.8\textwidth]{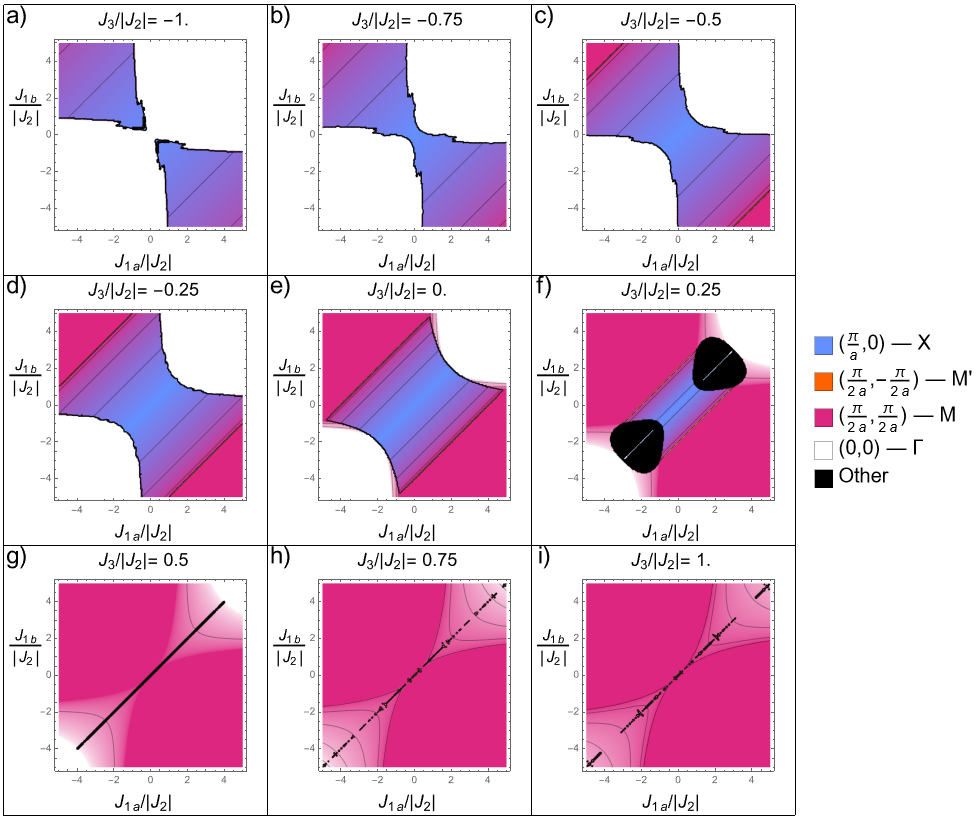}
	\caption{A density plot with colors corresponding to an ordering wavevector at various high-symmetry points in the Brillouin zone. Intermediate shades of those colors represent a minimum along the line between the two relevant points. $J_{2a}=J_{2b}\equiv J_2>0$ as in Fig.\ \ref{fig:classPhaseSpace}, though the axes cover a different range. Black indicates a minimum that is not on any of the high-symmetry lines. Thin black lines are contour lines showing constant ordering vector within an incommensurate phase. \label{fig:nearDensity}}
\end{figure}

Fig.\ \ref{fig:nearDensity} shows a density plot with non-AFM3 regions classified according to their magnetic ordering wavevector. We see that there exist regions of parameter space corresponding to crossing incommensurate phases. We observe from the figure that the phase with ordering wavevector which evolves from $\Gamma$ to $M$ smoothly is accessible in-plane for $J_3\geqslant 0$, occupying a larger region of the parameter space plane as $J_3$ increases. Paths through the full five-dimensional parameter space could also clearly cross such phases.  See section \ref{app:nearAFM3} of the Supplemental Material for additional details on incommensurate phases with smoothly varying ordering wavevector.

\section{Spin-Wave Fits to Inelastic Neutron Scattering Data \label{sec:SpinWaves}}

\begin{figure}[h]
	\includegraphics[width=0.7\textwidth]{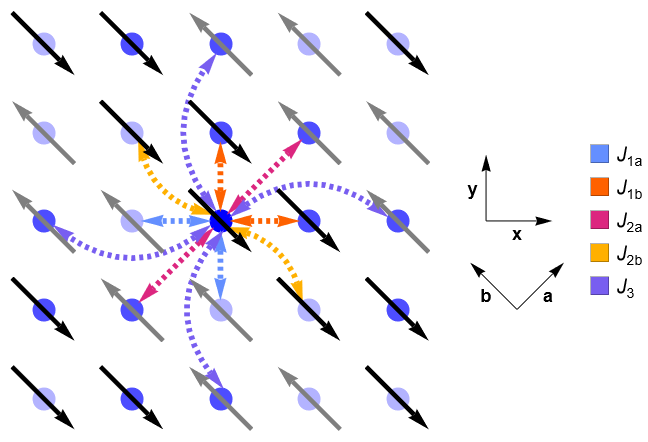}
	\caption{AFM3 ordering and coupling convention in FeTe. Notice that $a$-type couplings link anti-aligned spins and $b$-type couplings link aligned spins. $J_3$ couplings link anti-aligned spins in all directions. Additionally note the $\frac{\pi}{4}$ rotation between the lattice coordinate system $(x,y)$ and the magnetic coordinate system $(a,b)$.  \label{fig:lipscombeAFM3}}
\end{figure}

\begin{figure}[h]
	\includegraphics[width=0.48\textwidth]{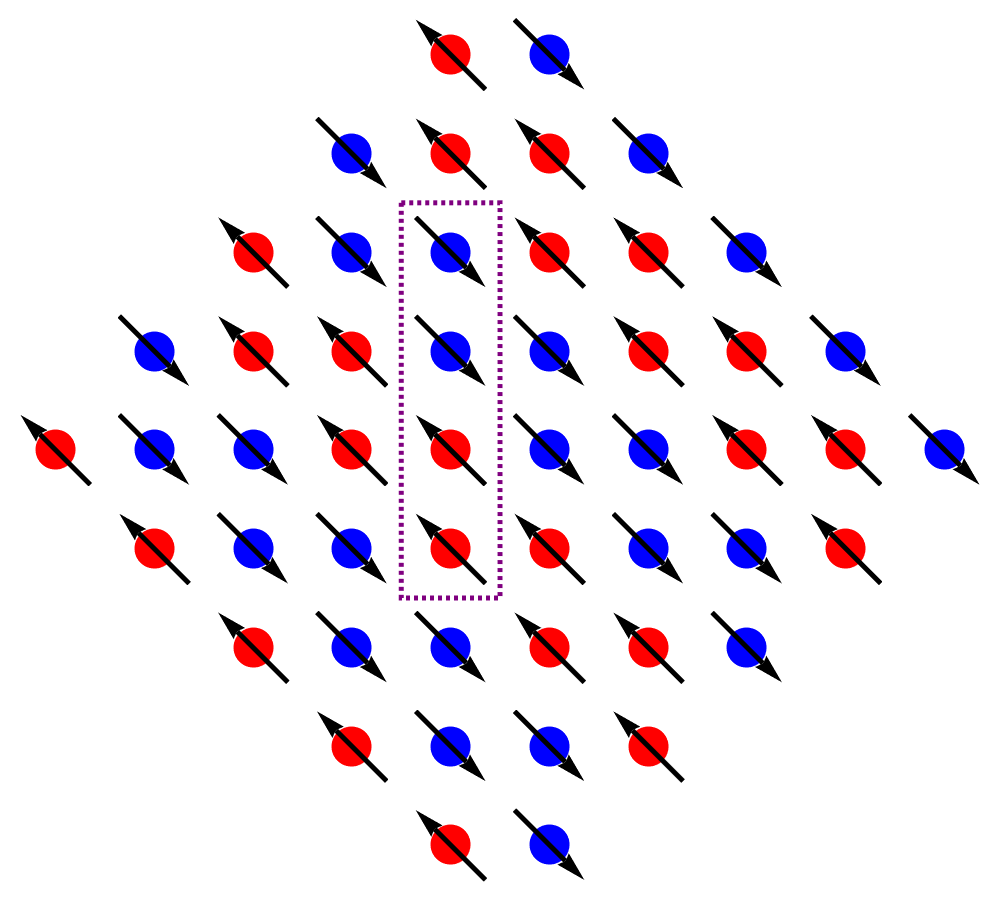}
	\caption{The magnetic unit cell.\label{fig:Quantcell}}
\end{figure}

The spin-wave dispersion of the Fang-Bernevig-Hu model for FeTe is obtained by applying a Holstein-Primakoff transformation \cite{auerbach1998interacting} --- which assumes small fluctuations about the AFM3 ground state --- to the Hamiltonian, (\ref{eq:FBHHamiltonian}). In the structural unit cell, Fig.\ \ref{fig:classcell}, $J_{1a}$ and $J_{1b}$ could be exchanged and this would equate to swapping the labeling of the $s$ sites and $t$ sites. In order to expand about a particular magnetic ground state, we must make a fixed choice of unit cell which amounts to specifying which of $J_{1a}$ and $J_{1b}$ links aligned sites and which links anti-aligned sites. We adopt the convention of Lipscombe \textit{et al}.\ as shown in Fig.\ \ref{fig:lipscombeAFM3}, and as such will enforce $J_{1a}>J_{1b}$ on our solutions. The unit cell chosen is displayed in Fig.\ \ref{fig:Quantcell}. For details of the calculation of the dispersion and intensity see section \ref{app:matReduction} of the Supplemental Material.

\begin{figure}[ph]
	\includegraphics[width=0.5\textwidth]{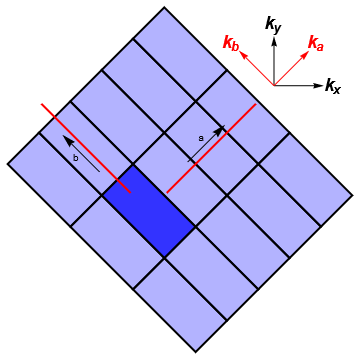}
	\caption{The cuts taken through the dispersion in reciprocal lattice space to match the Lipscombe \textit{et al}.\ data.  The darker zone is the first Brillouin zone.  \label{fig:BZCuts}}
\end{figure}

\begin{figure}[ph]
\hspace*{2cm}
	\includegraphics[width=0.75\textwidth]{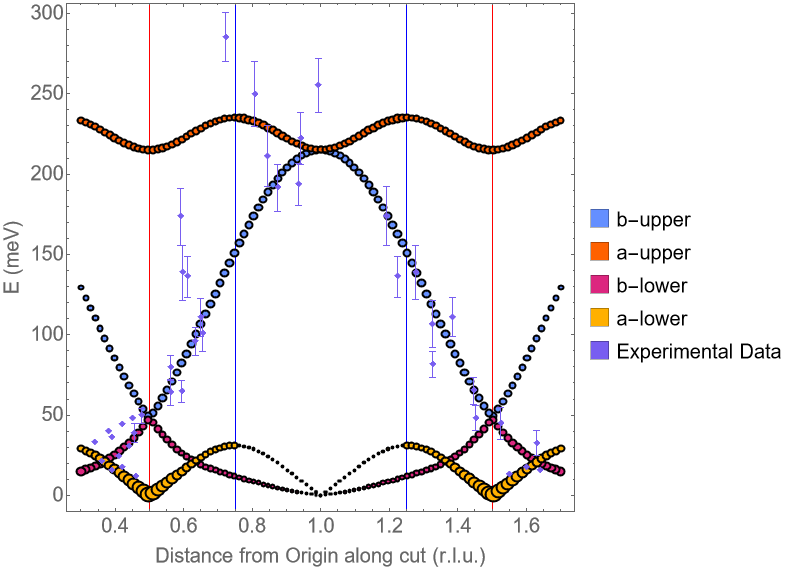}
	\caption{A plot produced from our spin wave calculations using the parameter values and data presented by Lipscombe \textit{et al}. in their figure 4(a) \cite{lipscombe2011spin}: $J_{1a}=-17.5$, $J_{1b}=-51.0$, $J_{2a}=21.7$, $J_{2b}=21.7$, and $J_{3}=6.8$.   \label{fig:lipscombeValues}}
\end{figure}

There turn out to be two distinct branches of the dispersion; however, the crystal samples studied are twinned and as such measurements are effectively probing two perpendicular directions of the crystal simultaneously. Therefore we observe a total of four separate branches in the dispersion plots from the combination of two branches in two separate directions. The cuts in reciprocal space are shown in Fig.\ \ref{fig:BZCuts} and are along the high-symmetry directions of the magnetic structure --- not the coordinate axes $x$ and $y$. These high-symmetry directions are labeled $a$ and $b$ to remain consistent with the Lipscombe \textit{et al}.\ analysis. The data presented by Lipscombe \textit{et al}.\ includes a correction for the form factor of the system and so we have also applied this correction to our predicted spin-wave dispersions. Fig.\ \ref{fig:lipscombeValues} shows the plot generated from our calculations using the parameter values provided in Fig.\ 4(a) of the Lipscombe \textit{et al}. paper \cite{lipscombe2011spin}.

\clearpage

Candidate sets of coupling constants were determined by minimizing a cost function based on the distance between experimental data points and the theoretically predicted spin-wave dispersion curves for a given choice of the coupling constants: $J_{1a}$, $J_{1b}$, $J_{2a}$, $J_{2b}$, and $J_3$. 
There are two ways in which a predicted spin-wave dispersion can fail to match the experimental data: it can predict a band in a region of the energy-momentum plane where there are no experimental data points, or it can fail to predict a band in a region where experimental data points do exist. The former case --- which we might call ‘missing data points’ --- can occur for many reasons, and thus is a less serious concern than the latter, which we might call ‘missing band’. Accordingly we have used a cost function that does not penalize the missing data point case, but does penalize the missing band case. Both linear and quadratic summations of these distances were used to find data fits and the best results from both groups of fits selected by eye.

Let $\mathbf b_i$ be discrete points sampled from the calculated spin-wave dispersion curves. They are therefore directly determined by the parameter values $J_{1a}$, $J_{1b}$, $J_{2a}$, $J_{2b}$, and $J_3$. Each experimental data point is divided into $m$ sub-points which span its error bars. Let $\mathbf l_{i,j}$ denote the $j$\textsuperscript{th} sub-point of the $i$\textsuperscript{th} data point. The linear and quadratic cost functions, $C_1$ and $C_2$ respectively, are then:
\begin{equation}
C_k(J_{1a},J_{1b},J_{2a},J_{2b},J_3)=
\begin{cases}
50+e^{R/10} & R>0 \\
\sum_i \min(d_{i,1},d_{i,2}, \ldots,d_{i,m})^k & R=0,
\end{cases}
\label{eq:costFunction}
\end{equation}
\begin{equation}
d_{i,j}=\min(|\mathbf l_{i,j}-\mathbf b_1|,|\mathbf l_{i,j}-\mathbf b_2|,\ldots,|\mathbf l_{i,j}-\mathbf b_n|),
\label{eq:costDistance}
\end{equation}
\begin{equation}
R(J_{1a},J_{1b},J_{2a},J_{2b},J_3)=\sum_i^N |\text{Im}(\mathbf b_i)|.
\label{eq:R}
\end{equation}

The $N$ in (\ref{eq:R}) indicates a sum over all sampling points, whereas in (\ref{eq:costDistance}) only the $n$ points from the calculated dispersion which exceed a given intensity threshold are included. In effect, the cost function does not ``see'' points of the calculated dispersion curve which are low intensity. Based on inspection of the intensity distribution across many parameter combinations, we chose a threshold of half the median intensity of the points $\mathbf b_i$. It is difficult to rigorously constrain the parameters to the set which produce an ordering wavevector at $M$; therefore, we must consider the entire parameter space as valid and ensure a high cost for unsuitable parameter combinations. $R(J_{1a},J_{1b},J_{2a},J_{2b},J_3)$  measures the magnitude of the imaginary part of the dispersion to achieve this.

For $R>0$ the cost functions $C_k$ have a downward slope to direct the minimization algorithm toward less unstable parameter configurations. The ``50'' and ``10'' in the $R>0$ output of (\ref{eq:costFunction}) are arbitrarily chosen with the former far larger than the typical cost of a plausible fit. In (\ref{eq:costDistance}), $d_{i,j}$ is the minimum distance between a given sub-point and any of the dispersion sampling points. The vector norm on this space is defined to include a rescaling in the energy coordinate by the height of the experimental data, and rescaling in the momentum coordinate by the width of the experimental data. $C_k$ is the sum of these scaled distances. Squaring each term in the quadratic case assigns a greater cost to data fits with, for example, a single data point far from the calculated dispersion than a fit in which many data points are slightly off the calculated dispersion. 

\begin{figure}[ph]
\vspace{-1.75cm}
\hspace{1.5cm}
	\includegraphics[width=0.8\textwidth]{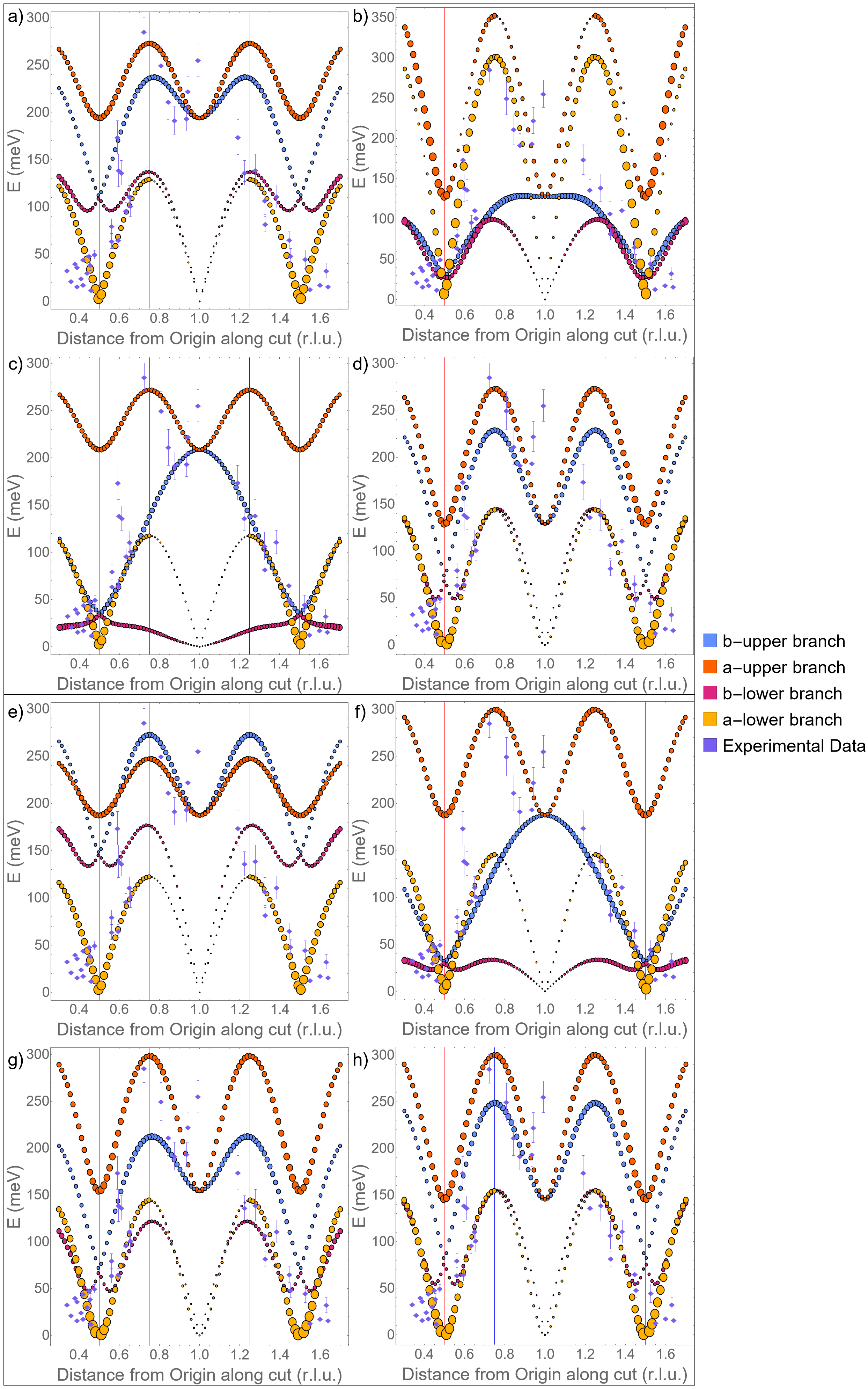}
	\caption{Possible spin-wave fits to inelastic neutron scattering data on FeTe with $J_{2a}=J_{2b}$ enforced. The radius of each point indicates relative intensity predicted by the model. Blue vertical lines (1.25 and 0.75) indicate a zone edge in the $a$ direction. Red vertical lines (0.5 and 1.5) indicate a zone edge in the $b$ direction; distance is measured in reciprocal lattice units in the $b$ direction. Parameter values for each plot are shown in Table \ref{tab:fixedVals}. Experimental data taken from ref.\ \cite{lipscombe2011spin}. \label{fig:fixedPlots}}
\end{figure}

\begin{figure}[ph]
\vspace{-1.75cm}
\hspace{1.5cm}
	\includegraphics[width=0.8\textwidth]{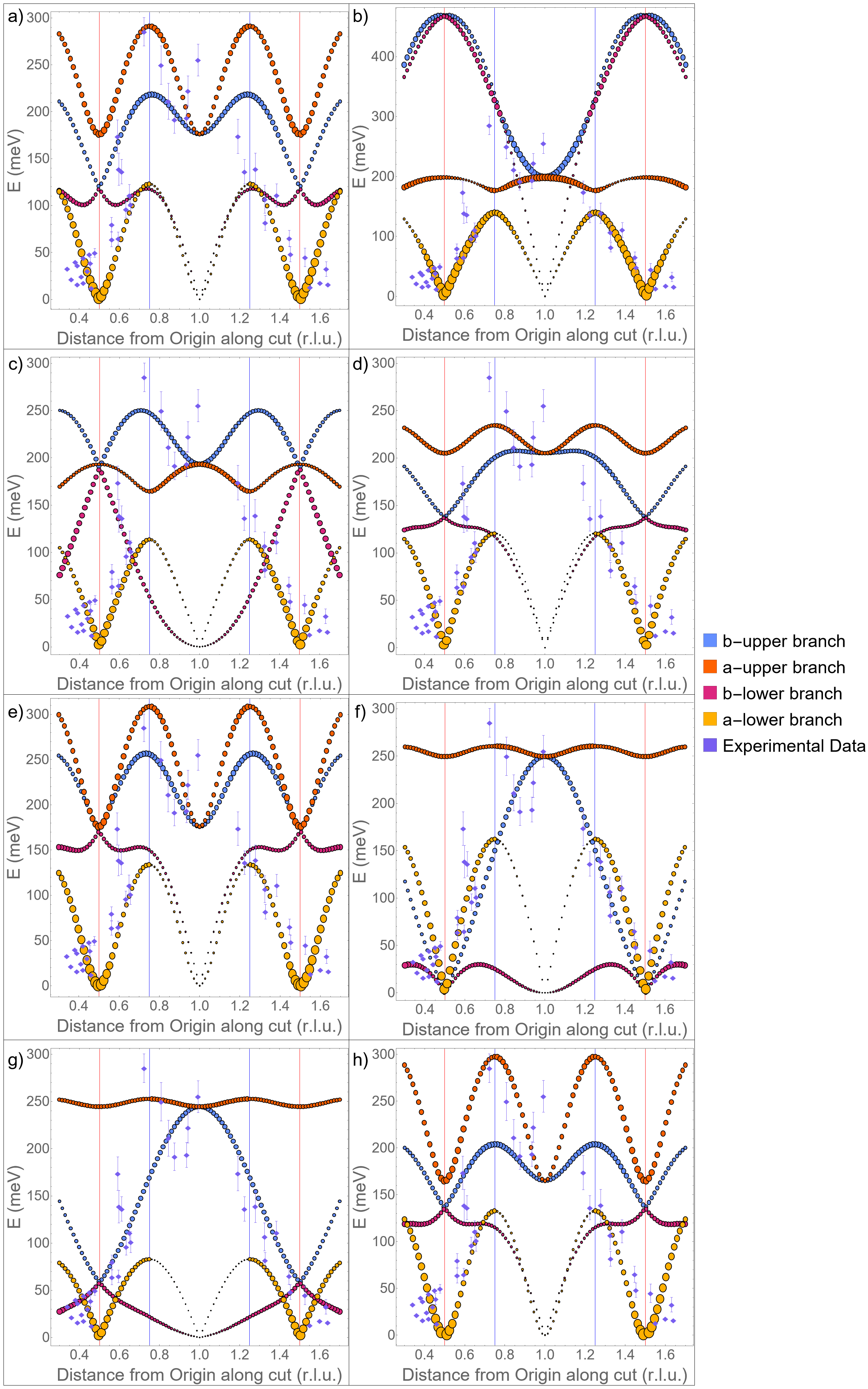}
	\caption{Possible spin-wave fits to inelastic neutron scattering data on FeTe with all parameters free. Notation the same as in Fig.\ \ref{fig:fixedPlots}. Parameter values for each of these plots are shown in Table \ref{tab:freeVals}.  Experimental data taken from ref.\ \cite{lipscombe2011spin}.  \label{fig:freePlots}}
\end{figure}

This minimization procedure has produced many results that are equally plausible fits to the data. Fig.\ \ref{fig:fixedPlots} shows some possibilities when we fix $J_{2a}=J_{2b}$. The parameter values for these plots are shown in Table \ref{tab:fixedVals}. Relaxing the constraint $J_{2a}=J_{2b}$ and allowing all five parameters to vary produces the plots shown in Fig.\ \ref{fig:freePlots}. The parameter values for these plots are shown in Table \ref{tab:freeVals}. In each of these plots we observe that the lower branches near the center have low intensity, indicated by the plot marker radius. This must be the case for any reasonable fits as for any parameters there will be theoretically predicted bands in the region, yet none appear in the data. Note that the marker radius has both a minimum and maximum set for clarity of presentation, and only relative intensity is displayed rather than an absolute scale.
\setlength{\tabcolsep}{1.5em}
\begin{table}[h]
\begin{tabular}{c|c|c|c|c|c}
Fig.\ \ref{fig:fixedPlots} plot label & $J_{1a}$ &$J_{1b}$&$J_{2a}$&$J_{2b}$&$J_3$ \\
\hline 
a &$-16.08$&$-36.03$&$7.65$&$7.65$&$37.08$  \\
\hline 
b & $19.83$ & $12.80$ & $73.24$ & $73.24$ & $42.14$ \\
\hline
c &$-14.53$&$-38.59$&$40.07$&$40.07$&$17.14$ \\
\hline 
d &$-24.34$ & $-32.24$ & $11.82$ & $11.82$ & $43.66$ \\
\hline 
e & $-9.96$ & $-31.22$& $-22.07$ &$-22.07$&$46.86$ \\
\hline
f & $-22.46$ & $-38.59$ & $49.03$ & $49.03$ & $24.16$ \\
\hline
g & $-28.12$ & $-38.55$ & $25.96$ & $25.96$ &$38.88$ \\
\hline 
h &$-27.21$ & $-36.40$ & $13.91$ & $13.91$ & $46.88$
\end{tabular}
\caption{Parameter values for the $J_{2a}=J_{2b}$ plots shown in Fig.\ \ref{fig:fixedPlots} in meV.\label{tab:fixedVals}}
\end{table}

\begin{table}[h]
\begin{tabular}{c|c|c|c|c|c}
Fig.\ \ref{fig:freePlots} plot label & $J_{1a}$ &$J_{1b}$&$J_{2a}$&$J_{2b}$&$J_3$ \\
\hline
a& $-27.67$&$-42.04$&$27.20$&$18.21$&$32.81$\\
\hline
b&$95.34$&$9.19$&$10.24$&$-55.32$&$13.73$\\
\hline
c&$62.88$&$-12.70$&$58.81$&$-21.78$&$-20.32$\\
\hline 
d&$1.73$&$-28.50$&$16.02$&$7.54$&$21.26$\\
\hline
e&$-30.41$&$-43.73$&$27.45$&$8.60$&$36.98$\\
\hline
f&$19.84$&$-24.52$&$26.06$&$48.23$&$18.57$\\
\hline
g&$4.56$&$-42.45$&$21.32$&$27.17$&$7.86$\\
\hline
h&$-29.15$&$-41.22$&$42.04$&$20.90$&$28.68$
\end{tabular}
\caption{Parameter values for the plots shown in Fig.\ \ref{fig:freePlots} in meV. \label{tab:freeVals}}
\end{table}

\pagebreak

\section{Discussion \label{sec:Discuss}}
\subsection{Analysis of Spin Wave Data Fits}
We now consider the implications of our results from sections \ref{sec:ClassicalPhase} and \ref{sec:SpinWaves}. We begin with our spin wave data fits and work backwards. Each of the plots in Fig.\ \ref{fig:fixedPlots} and Fig.\ \ref{fig:freePlots} has both desirable and undesirable features. Some, such as Figs.\ \ref{fig:freePlots}(f) and \ref{fig:fixedPlots}(f), provide better fits to the low energy data than others, yet high energy data remain with no overlap between the experimental error bars and the calculated spin-wave dispersion. A survey of all plots presented suggests many ways in which the parameters may be chosen to provide equally satisfactory fits to the data, each with different drawbacks. We see that a) no plot fits all data points within the error bars and b) most plots have moderate or high intensity regions of the bands at low or moderate energy which are not observed in the data. While one may suggest experimental justification for some of these absences, particularly at high energies or for largely flat bands, it does not seem likely that all such absences can be justified for any given plot. 

To give a concrete example, consider Fig.\ \ref{fig:freePlots}(f). This fit seems particularly good at capturing the low energy ($< 50$ meV) data points and is a rather good fit up to $\sim150$ meV. While the uppermost branch, which is a cut in the $a$ direction, is not reflected in the data this can perhaps be excused due to the non-dispersive nature of the branch. Despite these positive features, the fit remains imperfect. Above 175 meV, many of the data points do not lie on predicted bands within their error range. Furthermore, one might wonder why the lower branch in the $b$ direction is observed clearly outside of the zone these cuts are centered on (i.e. $<0.5$ and $>1.5$ r.l.u.) but not inside this range even though the band is dispersive and the same intensity is predicted. Similar observations of unsatisfactory properties can be made of any of the fits produced. In fact, almost entirely the same concerns arise upon inspection of the Lipscombe \textit{et al}.\ fit shown in Fig.\ \ref{fig:lipscombeValues}, though with emphasis on lack of observing the $a$-lower branch in the above specified range.

Having seen that there are many fits possible, note how broad the range of coupling constants is across such fits. Some have nearly isotropic $J_1$ values, such as Figs.\ \ref{fig:fixedPlots}(b), \ref{fig:fixedPlots}(d), \ref{fig:fixedPlots}(g), and \ref{fig:freePlots}(e). Others --- Figs.\ \ref{fig:freePlots}(c), \ref{fig:freePlots}(d), \ref{fig:freePlots}(f) --- are very anisotropic. Similar variation is seen across Fig.\ \ref{fig:freePlots} for $J_{2a}$ and $J_{2b}$. $J_3$ values vary from negative (Fig.\ \ref{fig:freePlots}(c)) to small and positive (Fig.\ \ref{fig:freePlots}(g)) to large and positive --- $\sim$ 50 meV and more than twice $J_2$ in Figs.\ \ref{fig:fixedPlots}(e) and \ref{fig:fixedPlots}(h).

The presence of many fits to the data which a) appear approximately equally suitable by eye, b) cover a broad range of parameter values, c) produce quantitatively very different dispersion cuts, and d) all have some number of unsatisfactory features suggest that the model applied to the system may not suitably describe it. At the very least, additional information on the nature of the system is required to constrain further the plausible regions of parameter space. Furthermore, we note a slight preference in the cost function minimisation toward $J_3$ values with magnitude either comparable to or greater than the other parameters. Given that $J_3$ coupling involves twice as many electron hops as other couplings, one may worry about the suitability of any derivation of the Heisenberg model based on an expansion around the atomic limit; and if an atomic-limit derivation of the Hamiltonian is not applicable, what would be the justification for setting couplings of longer range than $J_3$ to zero?

\subsection{Proximity to Incommensurate Phases}

\begin{figure}[pth!]
\hspace*{1cm}
	\includegraphics[width=0.95\textwidth]{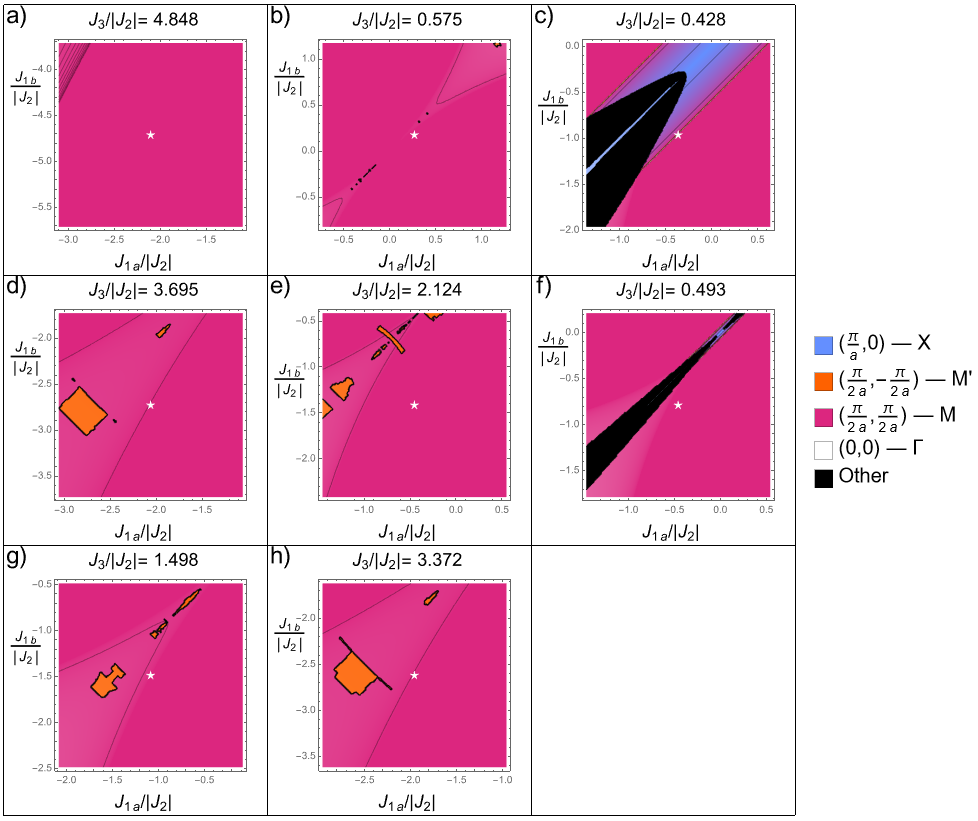}
	\caption{$J_{2a}=J_{2b}\equiv J_2$ best fits of inelastic neutron scattering data to calculated spin-wave dispersion curves shown in parameter space. Note the differing axis scales between plots. Parameter combinations corresponding to data fits in Fig.\ \ref{fig:fixedPlots} are indicated by white stars in the $J_{1a}-J_{1b}$ plane, and should not be confused for a region with structure factor minimum at the origin. \label{fig:fixedDensityClose}}
\end{figure}

\begin{figure}[pth!]
\hspace*{1cm}
	\includegraphics[width=0.95\textwidth]{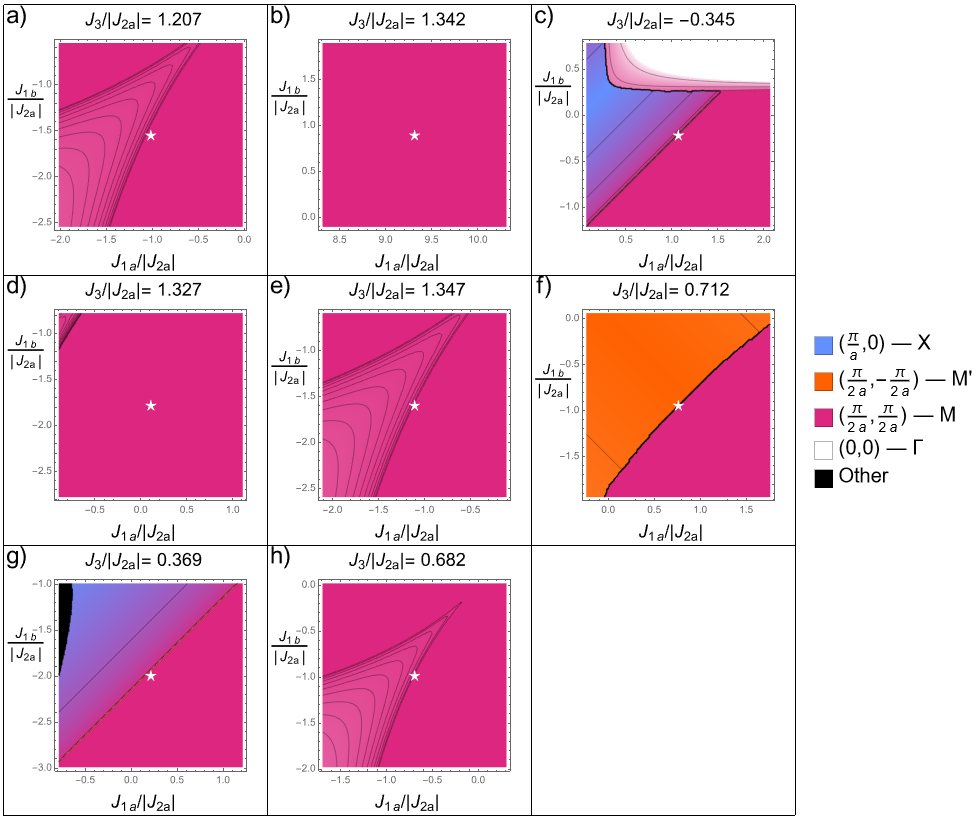}
	\caption{Best fits of inelastic neutron scattering data to calculated spin-wave dispersion curves with five free parameters shown in parameter space. Notation is the same as in Fig.\ \ref{fig:fixedDensityClose}. These points correspond the the data fits in Fig.\ \ref{fig:freePlots}. \label{fig:freeDensityClose}}
\end{figure}

Having seen our spin wave fits to the data, we may now return to discussion of the possibility that the FeTe sample is not in fact in the AFM3 phase. This suggestion is motivated further upon noticing that many of the fits use combinations of coupling constants that lie extremely close to phase boundaries in the $J_{1a}-J_{1b}$ plane.  Figs.\ \ref{fig:fixedDensityClose} and \ref{fig:freeDensityClose} show this proximity for our data fits. The Lipscombe \textit{et al}.\ fit \cite{lipscombe2011spin} shown in Fig.\ \ref{fig:lipscombeValues}, as well as several other plots shown in the original paper, also have parameter values which lie extremely close to phase boundaries. 

Intuitively one expects high-symmetry points to be preferred over others in the absence of information to the contrary. Manganese silicide provides a case which advises caution in exercising such intuition. MnSi was thought to be ferromagnetic \cite{bloch1975high, shinoda1966magnetic} until it was revealed that the magnetic ordering vector was slightly off the origin, leading to a helical spin pattern with a period of 180\,\AA \   \cite{ishikawa1976helical,ishikawa1977magnetic}. 

Furthermore, it has been observed that FeTe$_{1-x}$Se$_x$ has spin excitations with an intensity peak which progresses from $M$ towards $X$ as the energy is increased \cite{lumsden2010evolution,argyriou2010incommensurate}. Of particular note is the fact that the minimum in energy of these excitations appears to be at an incommensurate wavevector, which suggests that as the doping is reduced the long-range magnetic order may emerge with an incommensurate ordering wavevector rather than one at $M$. Intensity peaks near $X$ have also been observed in inelastic neutron scattering experiments on superconducting Fe$_y$Te$_{1-x}$Se$_x$ \cite{xu2016thermal}. 

As FeTe is doped with selenium the crystal first exhibits the emergence of a non-bulk superconducting phase, which occurs before long-range magnetic order is suppressed \cite{MARTINELLI2016phase}. Given the coexistence of incommensurate magnetism and superconductivity \cite{khasanov2009coexistence} at some values of doping, it seems reasonable to conclude that the superconducting mechanism is closely linked to the magnetic order of the system. In terms of the model, doping up to a certain threshold can likely be described by perturbing the coupling constants, reinforcing the idea that understanding these incommensurate phases with ordering wavevector which smoothly evolves away from $M$ is relevant to understanding the superconducting phase.

There have additionally been discussions of the magnetic order of iron telluride under excess iron doping leading to incommensurate helimagnetic order \cite{li2009first, fobes2014ferro,turner2009kinetic,bao2009tunable}, and high-pressure measurements resulting in destruction of magnetic order \cite{zhang2009density}. The former is particularly relevant to this data --- the sample measured with inelastic neutron scattering by Lipscombe \textit{et al}.\ is slightly doped with excess iron. This fact further favors the idea that the sample under measurement could be ordered at a wavevector slightly away from the AFM3 value. Moreover, the fact that small energy scales are relevant in determining the magnetic properties of FeTe may indicate that orbital effects could be significant. The degeneracy of some branches may lift under the action of spin-wave hybridization with an orbital degree of freedom \cite{ReichstetterPrepOrbital}; if this effect is substantial the calculated dispersion curves could be quite different.

The presence of short-range incommensurate magnetic order \cite{wen2009short, xu2018coexistence} in the same family of materials, FeTe$_{1-x}$Se$_x$, which enter superconducting phases as doping is increased \cite{MARTINELLI2016phase} further provides hope that understanding of the spin structure of FeTe and potential instabilities could lead to improved understanding of iron-based superconductors \cite{mook2010unusual}. While excess iron in the Lipscombe \textit{et al}.\ sample favors the $(\epsilon,\epsilon)$ incommensurate phase, a phase with ordering wavevector on the edge of the Brillouin zone is still plausible. In particular, we note that Figs.\ \ref{fig:fixedDensityClose}(c),  \ref{fig:freeDensityClose}(c), and \ref{fig:freeDensityClose}(g) show best fits on the boundary of this incommensurate phase. Regardless, it is entirely possible for the sample studied by Lipscombe \textit{et al}.\ to have a magnetic ordering wavevector offset from $M$ toward $\Gamma$ but evolve through $M$ and toward $X$ as Se doping occurs. 

We end the discussion of these near-AFM3 states by observing that, as FeTe appears susceptible to evolving an incommensurate ordering wavevector under mild doping, the continuous nature of such ordering wavevector evolution means that the material could be subject to deviations from pure AFM3 order throughout its spatial extent. In other words, it may be the case that the parameter $\epsilon$ which indicates the magnetic ordering could vary slightly throughout the material due to any small perturbation which might affect the coupling constants locally. Small variation in the ordering wavevector throughout the crystal --- as an indication of the susceptibility to a short-range magnetic phase under doping --- is one way in which the Fang-Bernevig-Hu model could be valid for the system, but applied to an exact AFM3 phase still fail to reproduce the inelastic neutron scattering data.

\subsection{Conclusion}
In this paper, we have motivated an interest in the magnetism observed in iron telluride in the context of the emergence of a superconducting phase, and explored an anisotropic magnetic model of the crystal developed by Fang, Bernevig, and Hu \cite{fang2009theory} --- which has couplings up to the third-nearest-neighbor. We have used this model to study the static magnetic ordering of the compound and highlighted the existence of incommensurate phases which have a smoothly varying magnetic ordering wavevector as they are traversed in parameter space. The model has also been applied to study the spin-wave dispersion curves for excitations around the AFM3 state. In the discussion of our results, we have shown that a great many combinations of coupling constants exist which provide equally suitable fits to the data. Such a broad parameter range, including relatively isotropic and anisotropic nearest-neighbor and next-nearest-neighbor couplings, suggest that we should not give significantly greater credence to any particular combination above others. We believe it is therefore reasonable to question whether the model we have applied to FeTe actually provides a suitable description of the system, or whether modifications are required.

The suggestion that the model might require modification is reinforced by several facts we have previously mentioned. First, cost function minimization has a tendency toward data fits with coupling constant values extremely close to a phase boundary. Second, FeTe is susceptible to developing an incommensurate ordering wavevector under doping by excess iron, and exhibits short-range incommensurate magnetic order under doping by selenium in place of tellurium. Third, many fits to the data have large (relative to other parameters) $J_3$, despite the terms beyond third-neighbor couplings being omitted from the model. If long-range interactions are significant then there is no reason a short-range model should describe the system accurately. Finally, the small energy scales separating neighboring phases mean that additional physics, such as orbital physics, could substantially change the prediction of the model by providing additional excitation channels.

We therefore suggest four possible interpretations:
\begin{enumerate}
	\item The most conservative possibility is that a short-range Heisenberg model is capable of describing the magnetic FeTe system, and it truly is in the AFM3 state, but we require analysis of the orbital physics to guide our hand in selecting appropriate coupling constants. Given that none of the fits obtained provide a truly satisfactory match to the data, we find this option the least likely.
	\item The Fang-Bernevig-Hu model is capable of describing FeTe, but the classical ground state ordering differs slightly from AFM3, resulting in a long range helix on top of the double stripe pattern. Equally, the ordering vector could vary throughout the material in space due to instability away from AFM3. To explore this possibility further, a spin wave analysis based around such a state would be required \cite{kataoka1987spin}. We have not conducted spin wave expansions about such incommensurate states, but believe it would be a useful endeavor for another paper.
	\item The Fang-Bernevig-Hu model may not be capable of capturing the essence of this system if orbital effects produce a qualitative difference in behavior. An orbital model in this case would provide a genuinely different behavior of this system rather than simply further constraining parameter space, while remaining in the atomic limit \cite{ReichstetterPrepOrbital}.
	\item Most radically, a local-moment model may be incapable of describing this system. If this is the case only an itinerant model would be capable of correctly fitting neutron scattering data, and long range interactions would be important.
\end{enumerate}

It is believed that the suppression of AFM3 order under doping with Se or S is intimately linked to the transition to a superconducting phase \cite{ma2022superconductivity} and hence it is likely important to understand the evolution of the magnetic ordering wavevector in the FeTe system as the coupling parameters are perturbed. We conclude that an improved description of magnetism in the FeTe system is required to understand the origin of the superconducting phase; however, it is clear that much additional work is required in order to describe the rich physics at play.

\clearpage
\appendix
\begin{center}
{\Large Supplemental material for ``Magnetic order and magnetic excitations in FeTe:\ how good is the short-range Heisenberg model?''}
\end{center}

\section{Anisotropic and Ferromagnetic $J_2$ Parameter Spaces \label{app:minusJ2}}

\begin{figure}[pth]
\hspace*{1cm}
	\includegraphics[width=0.65\textwidth]{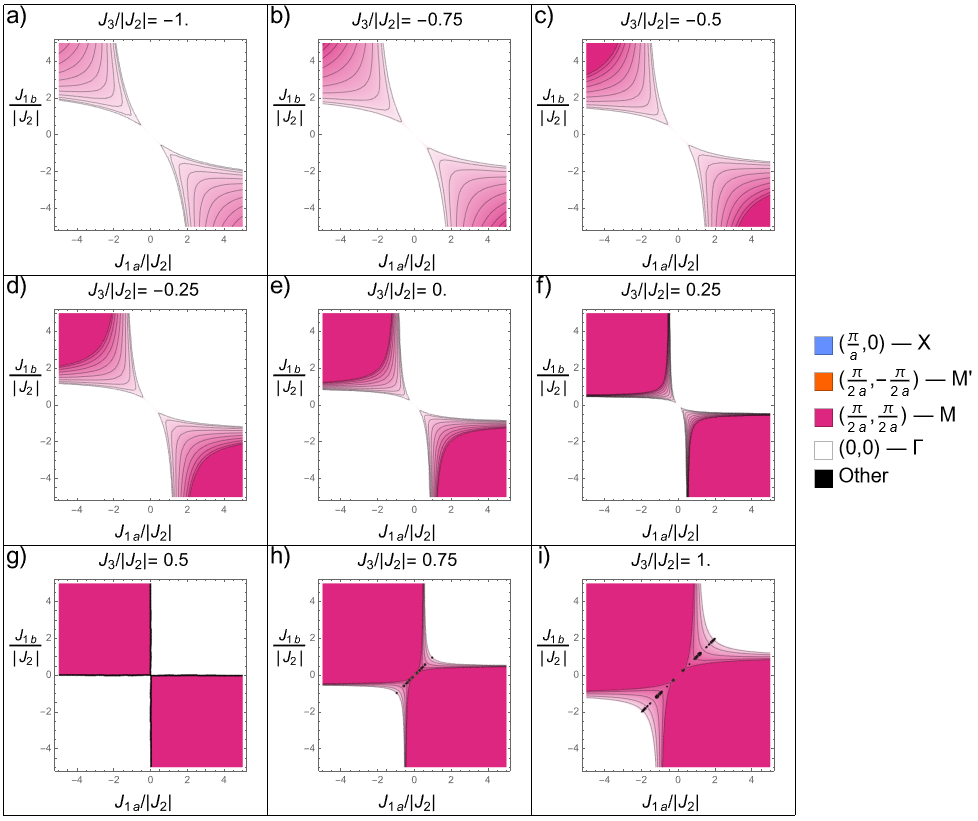}
	\caption{Parameter space plane plots showing the static magnetic ordering wavevector with $J_{2a}=J_{2b}\equiv J_2<0$. \label{fig:PhaseSpaceMinus}}
\end{figure}

\begin{figure}[pbh]
\hspace*{1cm}
	\includegraphics[width=0.65\textwidth]{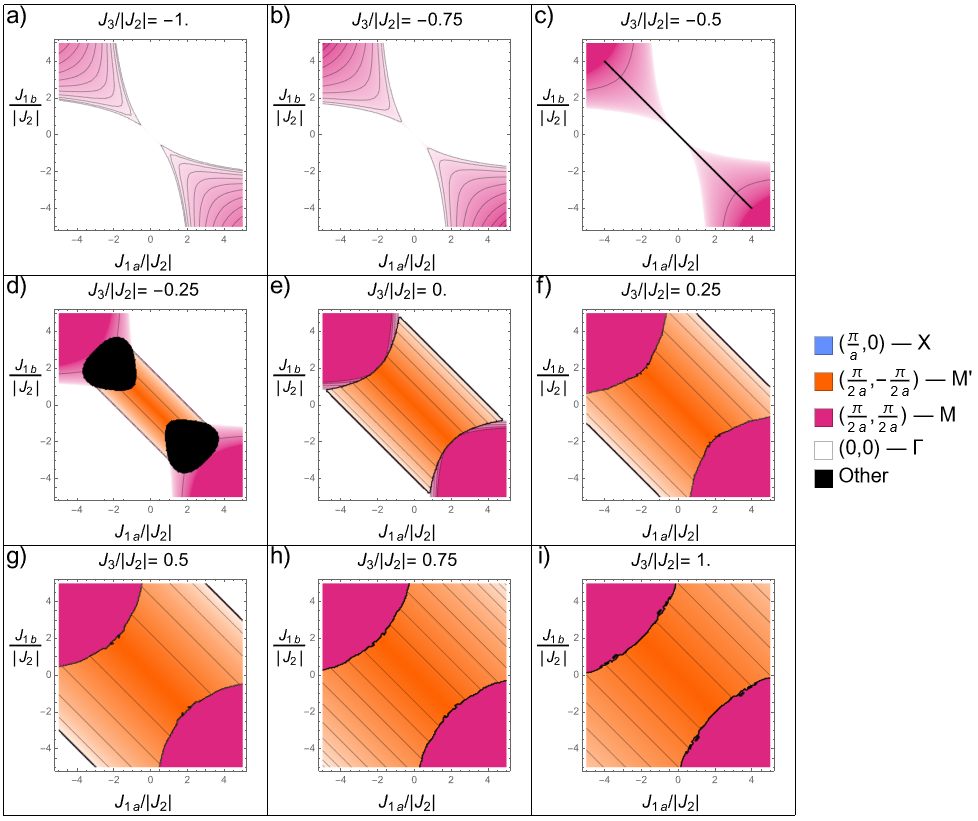} 
	\caption{ Parameter space plane plots showing the static magnetic ordering wavevector with $J_{2a}=-J_{2b}\equiv J_2<0$. \label{fig:PhaseSpaceMinusJ2a}}
\end{figure}

\begin{figure}[pth]
\hspace*{1cm}
	\includegraphics[width=0.65\textwidth]{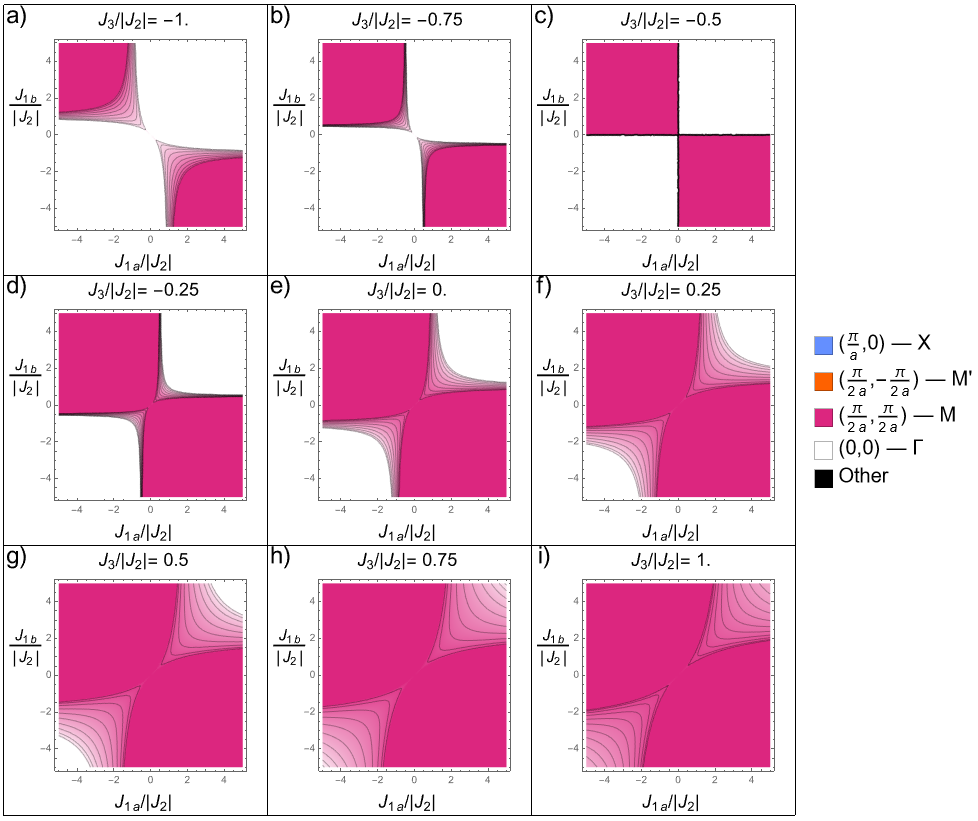}
	\caption{Parameter space plane plots showing the static magnetic ordering wavevector with  $J_{2b}=-J_{2a}\equiv J_2<0$. \label{fig:PhaseSpaceMinusJ2b}}
\end{figure}

Fig.\ \ref{fig:nearDensity} shows the $\frac{J_{1a}}{J_2}-\frac{J_{1b}}{J_2}$ parameter space for isotropic and positive $J_{2}$ but we may also consider how it changes for other $J_{2a}$ and $J_{2b}$ combinations. Fig.\ \ref{fig:PhaseSpaceMinus}, Fig.\ \ref{fig:PhaseSpaceMinusJ2a}, and Fig.\ \ref{fig:PhaseSpaceMinusJ2b} show plots for $J_{2a}$ and $J_{2b}$ negative, $J_{2a}$ negative, and $J_{2b}$ negative respectively. While the specific shape of the AFM3 region changes between these different slices of phase space, the general features remain the same.  One qualitative difference worth mentioning is the opening up of the central region of the plane in Fig.\ \ref{fig:PhaseSpaceMinusJ2b} where $J_{2b}<0$ and $J_3/|J_2|>0$. It would seem one can get arbitrarily close to the $J_{1a}=J_{1b}$ line from either side while remaining in the AFM3 state in this case. Just as there is a trade off between $J_{1a}$ and $J_{1b}$ isotropy and keeping $J_3$ small, there appears to be a similar trade off allowing us to sacrifice $J_{2a}$ and $J_{2b}$ isotropy instead. We additionally note that negative $J_2$ suppresses states with ordering wavevector along the edge of the Brillouin zone except AFM3 itself.

\begin{figure}[pbh]
\hspace*{1cm}
	\includegraphics[width=0.65\textwidth]{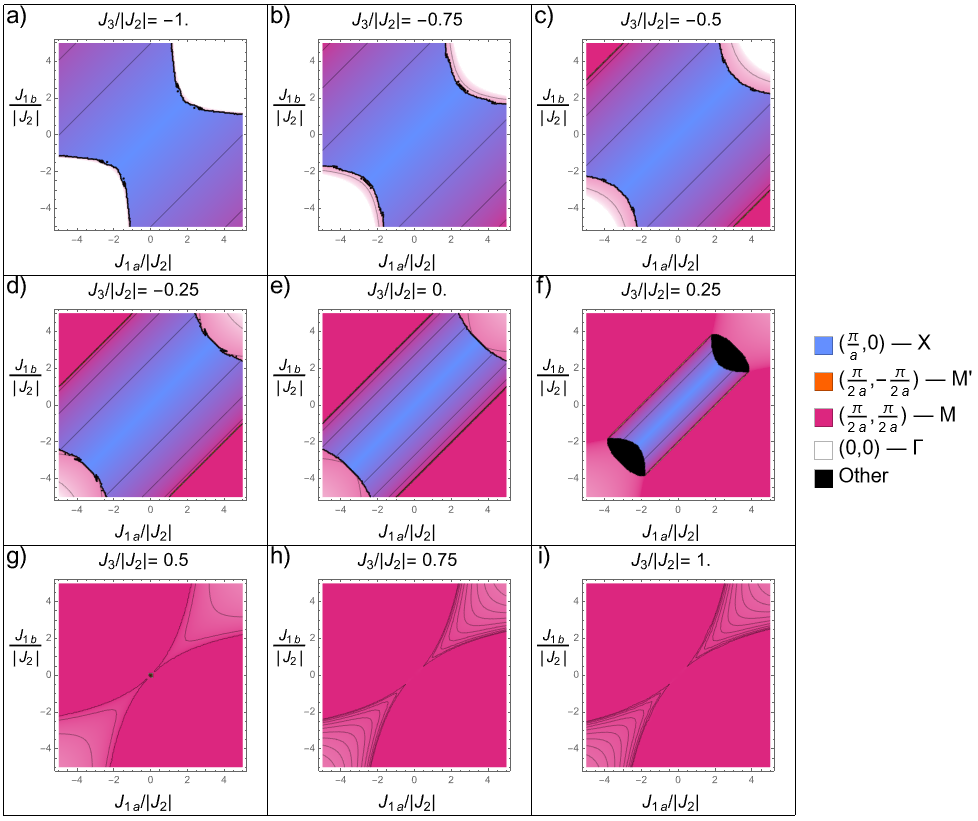}
	\caption{Parameter space plane plots showing the static magnetic ordering wavevector with $J_{2a}=3J_{2b}>0$ and $J_{2b}\equiv J_2>0$. \label{fig:PhaseSpaceMagnitudeJ2a}}
\end{figure}

\begin{figure}[h]
\hspace*{1cm}
	\includegraphics[width=0.65\textwidth]{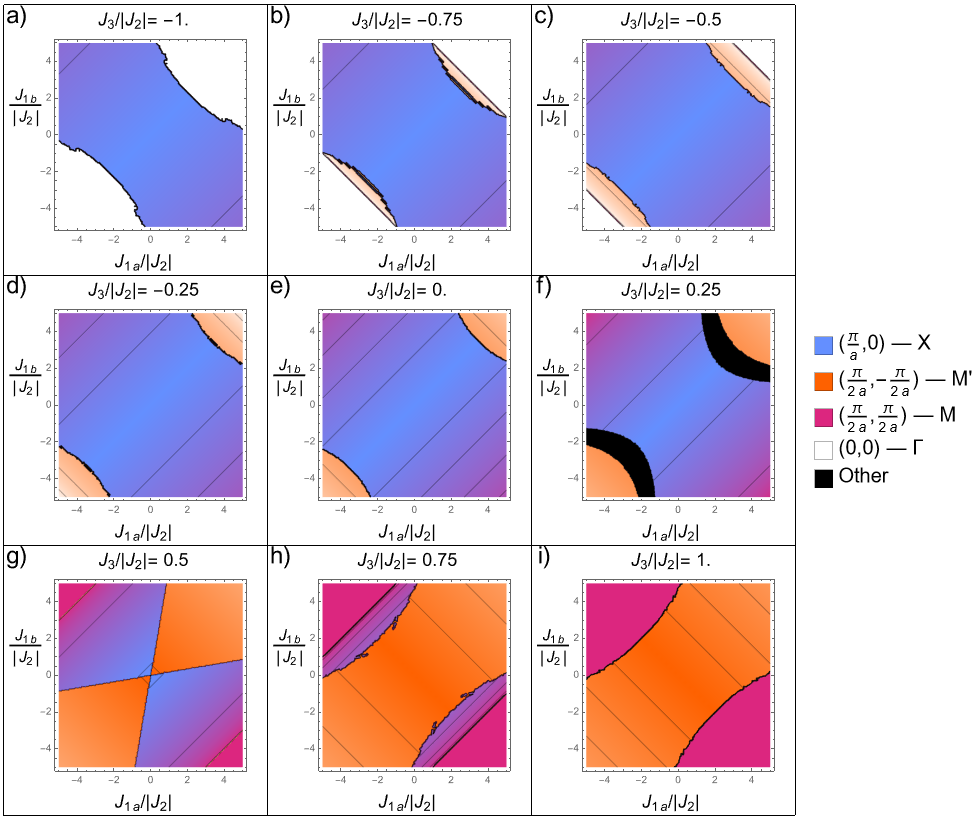}
	\caption{Parameter space plane plots showing the static magnetic ordering wavevector with $J_{2b}=3J_{2a}>0$ and $J_{2a}\equiv J_2$. \label{fig:PhaseSpaceMagnitudeJ2b}}
\end{figure}

Figs.\ \ref{fig:PhaseSpaceMagnitudeJ2a} and \ref{fig:PhaseSpaceMagnitudeJ2b} show the plane for positive $J_{2a}$ and $J_{2b}$ with different magnitudes where we observe an opening of the central region in the former and the appearance of $M'$ minima in the latter. The difference between Fig.\ \ref{fig:PhaseSpaceMinusJ2a} and Fig.\ \ref{fig:PhaseSpaceMagnitudeJ2b} is that in the latter we have both $X$ and $M'$ minima and $J_3$ determines which is dominant in the plane. We can understand why $M'$ regions appear by comparing their energy with that of the AFM3 ordering. For $M'$ to have lower energy than $M$ we require $|J_{1a}-J_{1b}|<2(J_{2b}-J_{2a})$ which is only possible for $J_{2a}<J_{2b}$. Note that this is a necessary but not sufficient condition for $M'$ to be a global minimum of the structure factor.


\pagebreak

\section{Other Incommensurate Phases \label{app:nearAFM3}}

\begin{figure}[pht]
	\includegraphics[width=0.5\textwidth]{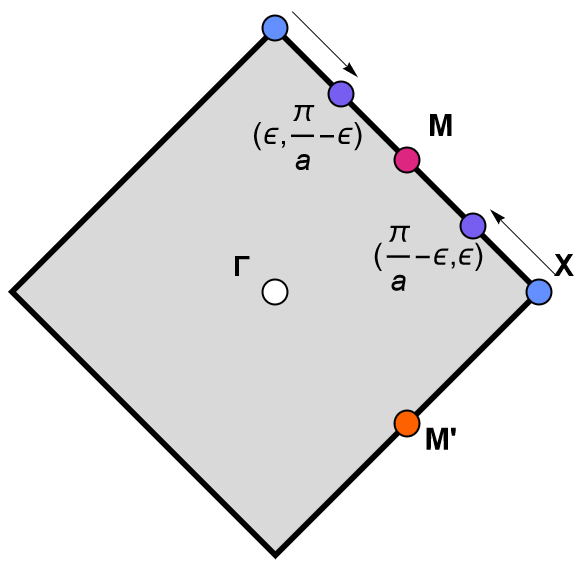}
	\caption{The structural Brillouin zone with relevant points shown for the incommensurate phase with ordering wavevector along the zone edge. \label{fig:classBZEdge} }
\end{figure}

\begin{figure}[phb]
	\includegraphics[width=0.9\textwidth]{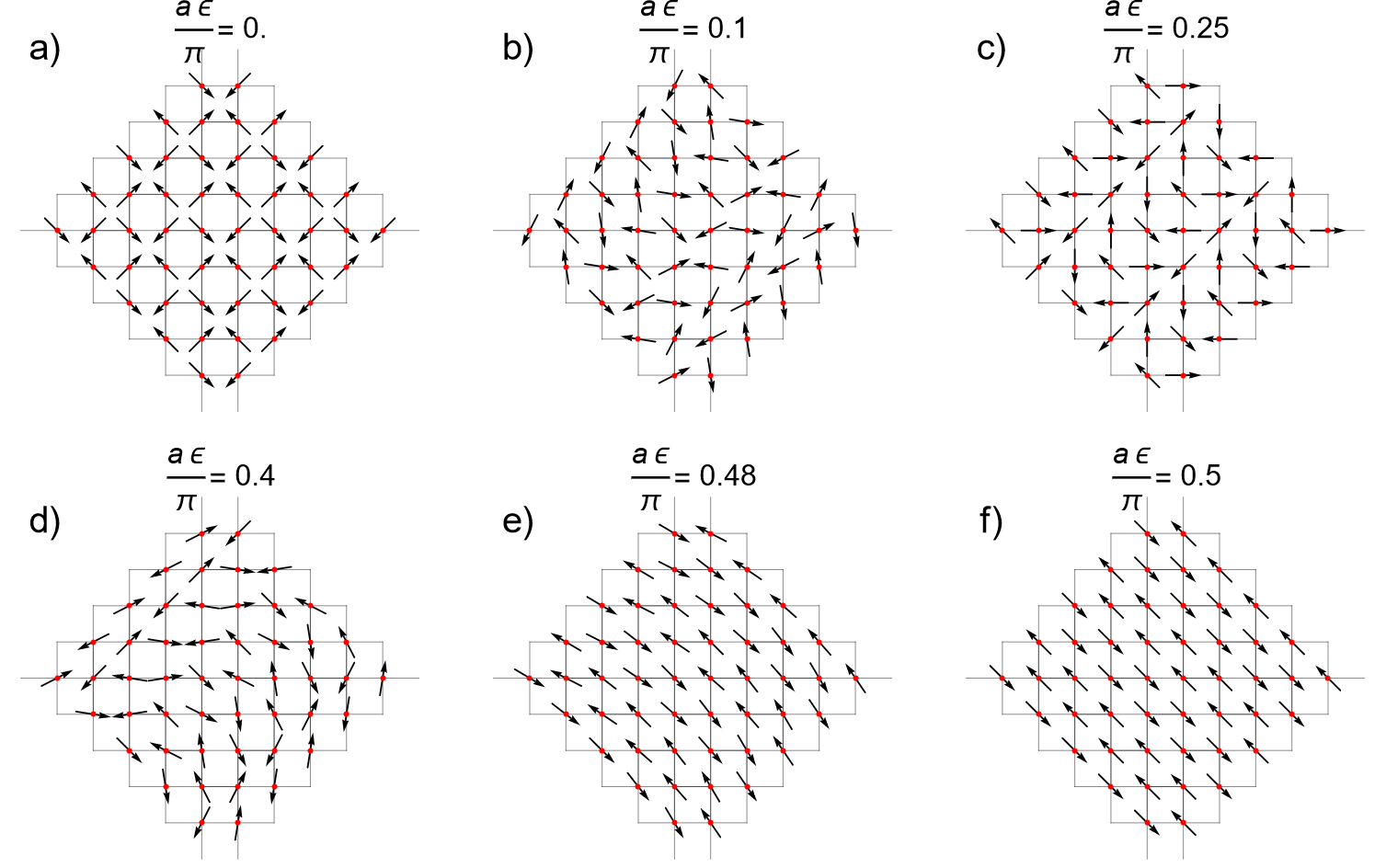}
	\caption{The real-space magnetic ordering resulting from a structure factor minimum at $(\frac{\pi}{a}-\epsilon,\epsilon)$. When $\epsilon=0$ the $t$ sites can be rotated freely with respect to $s$ sites with the same transformation applied in every unit cell. \label{fig:nearDensityPlots} }
\end{figure}

A minimum in the structure factor branches, (\ref{eq:gs}) and (\ref{eq:gt}) in the main text, determines the static ordering of the system. An extremal point exists at $(\frac{\pi}{a}-\epsilon,\epsilon)$ if
\begin{equation}
\sin(a \epsilon)=\frac{|J_{1a}-J_{1b}|}{4(J_{2b}-2 J_3)}.
\label{eq:nearAFM3}
\end{equation}
Fig.\ \ref{fig:classBZEdge} shows the positions of these minima in the structural Brillouin zone. Fig.\ \ref{fig:nearDensityPlots} shows the real-space magnetic ordering for minima along this line. As discussed in section \ref{ssec:AMF3Regions} of the main paper, when the structure factor branches touch the orientation of the two sites in the unit cell is independent; therefore, there exists a collection of continuously connected ground states at $\epsilon=0$ of which Fig.\ \ref{fig:nearDensityPlots}(a) is one example. Fig.\ \ref{fig:nearDensity} demonstrates that smooth evolutions of the ordering wavevector from $M$ to $X$ are actually realized in parameter space. In particular we observe that this incommensurate phase is more accessible for more negative $J_3$ which is why this phase is a less favorable candidate for the sample studied by Lipscombe \textit{et al}.. It should not be ruled out, however, as the phase still exists for $J_3<0.5 J_2$. No simple equation like (\ref{eq:nearAFM3}) exists for incommensurate phases with ordering wavevectors on the lines $\Gamma$ -- $M$, $\Gamma$ -- $X$, or $\Gamma$ -- $M'$.

\begin{figure}[ph]
\hspace*{1.5cm}
	\includegraphics[width=\textwidth]{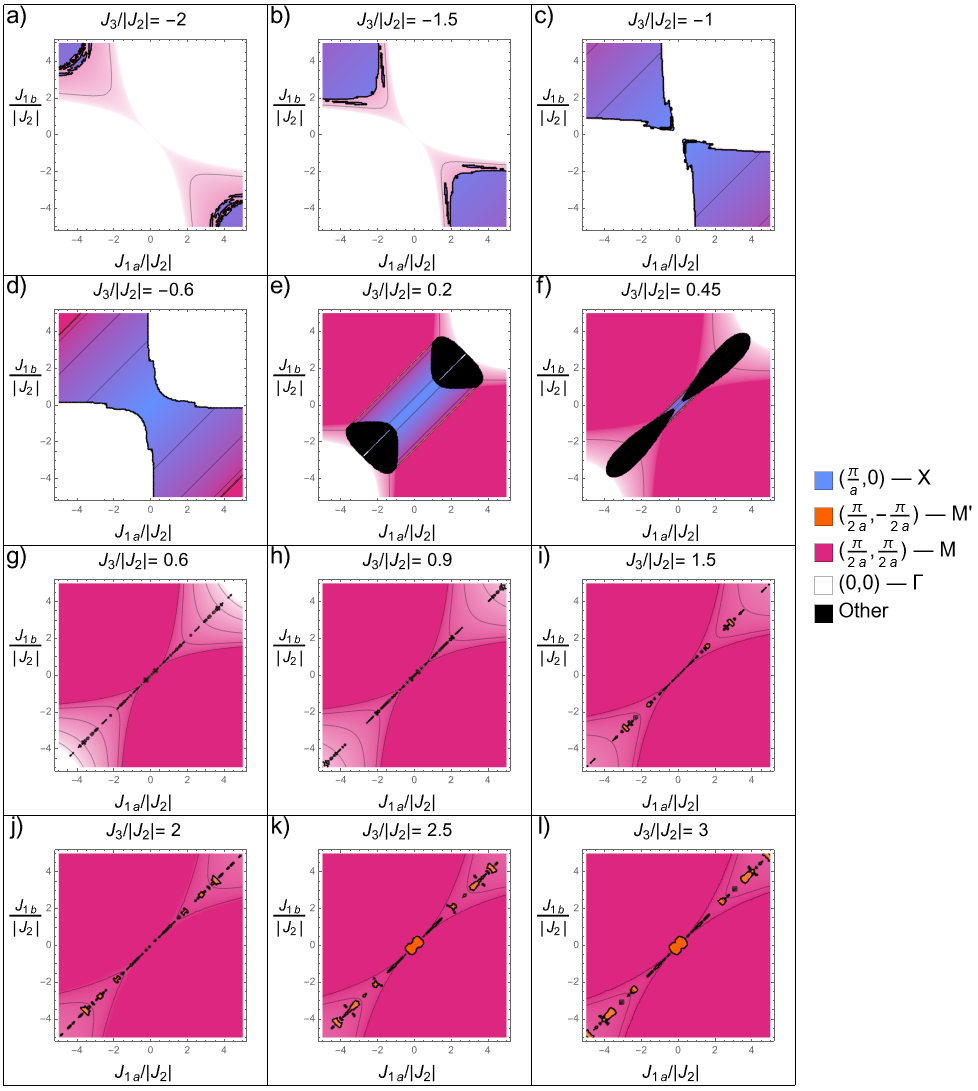}
	\caption{Plots in the $J_{1a}/|J_{2}| - J_{1b}/|J_{2}|$ parameter space plane indicating the static magnetic ordering wavevector for particular parameter combinations. These plots have $J_{2a}=J_{2b}\equiv J_2>0$ and demonstrate a broader range of $J_3/|J_2|$ than in the main text. \label{fig:moreJ3Range} }
\end{figure}

Fig.\ \ref{fig:moreJ3Range} shows a broader range of $J_3$ values in the $J_{1a}-J_{1b}$ plane for isotropic and equal $J_{2a}$ and $J_{2b}$ to demonstrate further the existence of other incommensurate phases throughout parameter space. We observe gradual shade changes indicative of incommensurate phases with ordering wavevector along lines between most pairs of high-symmetry points. Small regions appear to open up where $M'$ is a minimum for large $J_3$. As stated in section \ref{app:minusJ2} of the Supplemental Material, for the structure factor to have a lower value at $M'$ than $M$, $|J_{1a}-J_{1b}|<2(J_{2b}-J_{2a})$ with the right hand side being zero when $J_2$ isotropy is enforced. Hence we see these points in the Brillouin zone have closest energy when $J_{1a}$ and $J_{1b}$ are close which is precisely where we see these anomalies. We suggest these regions in the plot are therefore unphysical and likely due to numerical error.

\begin{figure}[ph]
\hspace*{1.5cm}
	\includegraphics[width=\textwidth]{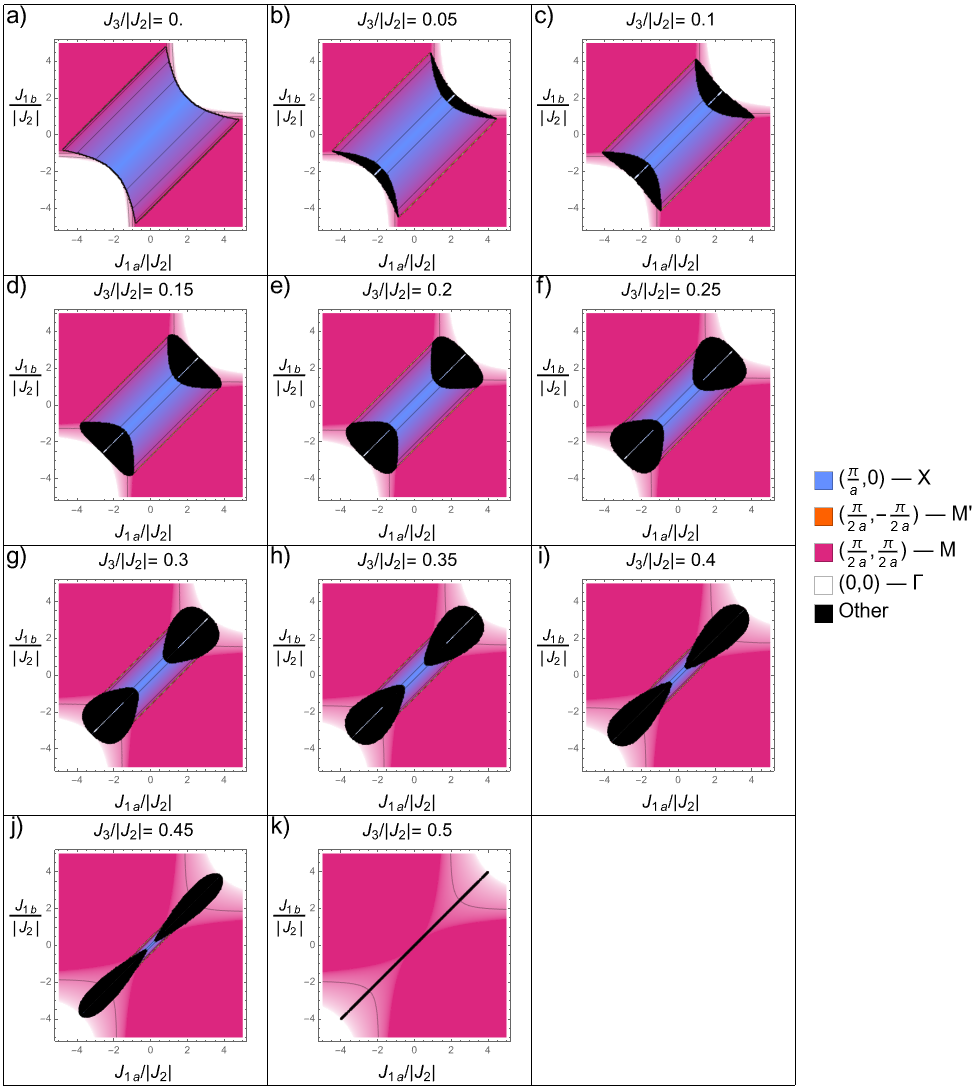}
	\caption{Evolution of the non-high-symmetry region (black) in the $J_{1a}/|J_{2}|-J_{1b}/|J_{2}|$ parameter space plane as $0\leqslant J_3/|J_2|\leqslant 0.5$ changes. These plots have $J_{2a}=J_{2b}\equiv J_2>0$.\label{fig:smallJ3Evol} }
\end{figure}

\begin{figure}[h!]
	\includegraphics[width=0.8\textwidth]{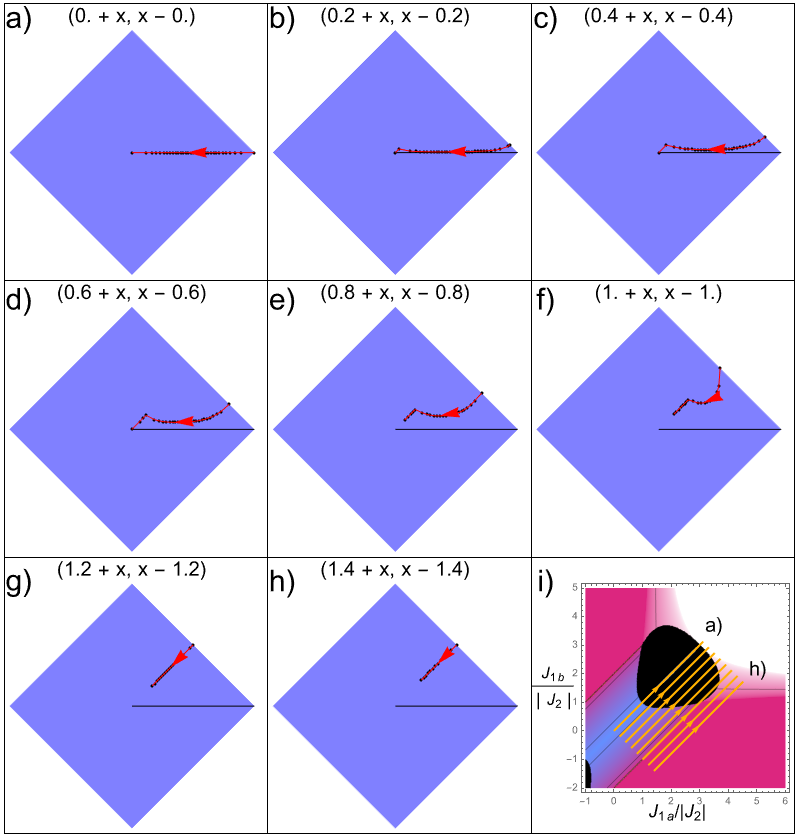}
	\caption{Evolution of the structure factor minimum upon passing through a non-high-symmetry region of parameter space. Black dots are the sampling points with a red line connecting them. The black line serves only to make it easier to observe that the ordering vector deviates from the high-symmetry line rapidly upon leaving $J_{1a}=J_{1b}$. $x$ parameterises the path taken in parameter space. Panel (i) shows the cuts along which this evolution is observed in the plane with $J_3=0.25\;J_2$ and isotropic $J_2$.  \label{fig:BlackRegionTransition} }
\end{figure}

A feature of the parameter space which we have not yet addressed is the prominent region which appears for $0<\frac{J_3}{J_2}<0.25$ corresponding to an ordering vector other than the high-symmetry points or lines between them. Fig.\ \ref{fig:smallJ3Evol} shows how this region changes in the relevant range. We may wonder whether this phase, which is certainly incommensurate, has an ordering wavevector which can smoothly approach AFM3. Fig.\ \ref{fig:BlackRegionTransition} shows how the structure factor minimum evolves as we move through this region for $J_3=0.25 J_2$.

\begin{figure}[h!]
	\includegraphics[width=0.8\textwidth]{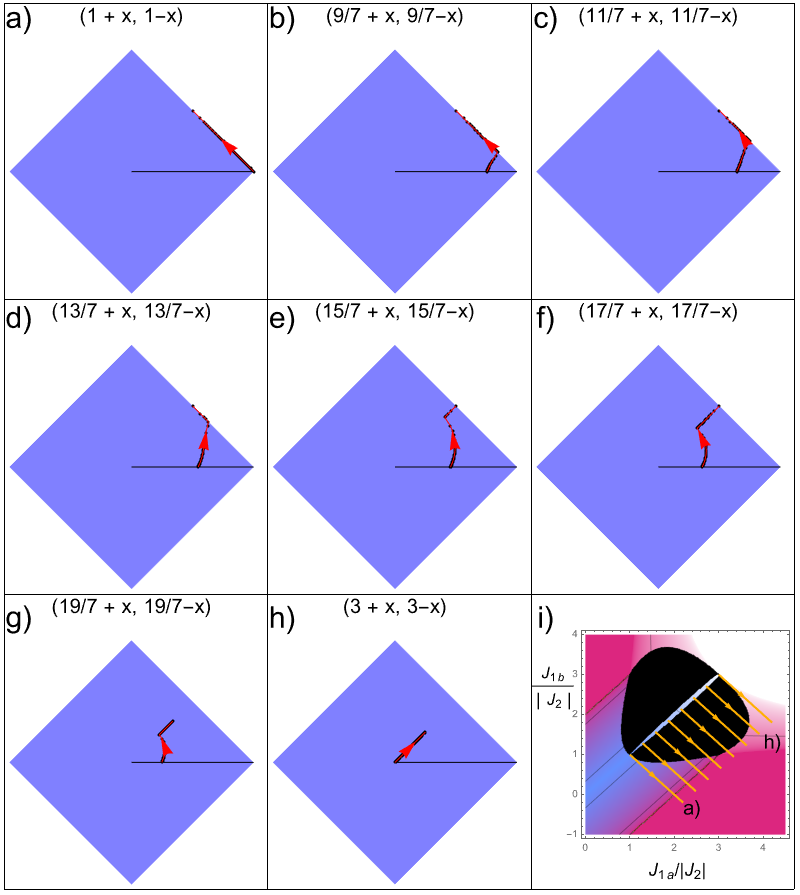}
	\caption{Evolution of the structure factor minimum along cuts through the black region perpendicular to those in Fig.\ \ref{fig:BlackRegionTransition}. Panel (i) shows the cuts taken with $J_3=0.25\;J_2$ and isotropic $J_2$. \label{fig:ShortBlackRegion} }
\end{figure}

Fig.\ \ref{fig:BlackRegionTransition}(i) shows the lines along which the evolution is shown in Figs.\ \ref{fig:BlackRegionTransition}(a-h). We see that the phase has ordering wavevector along the high-symmetry line only for precisely $J_{1a}=J_{1b}$ (Fig.\ \ref{fig:BlackRegionTransition}(a)) and shifting off the $J_{1a}=J_{1b}$ line in parameter space displaces the path in the Brillouin zone from the $\Gamma$ -- $X$ line. Of course, as we move the path in parameter space our starting point shifts from $X$ toward $M$. What is perhaps unexpected is that rather than the minimum going from the starting point to $X$ and then towards $\Gamma$, the path taken is a ``shortcut'' through the bulk of the Brillouin zone. Fig.\ \ref{fig:ShortBlackRegion} shows the same types of plot but in the perpendicular direction to those of Fig.\ \ref{fig:BlackRegionTransition}. 

While an interesting quirk, we note that in Fig.\ \ref{fig:BlackRegionTransition} the evolution of the wavevector away from AFM3 in all cases seems to be in the two directions already discussed. On the other hand, extrapolating between  Fig.\ \ref{fig:ShortBlackRegion}(d) and Fig.\ \ref{fig:ShortBlackRegion}(e) suggests that there may be some combinations of parameters for which other states exist arbitrarily close to AFM3. This is not the case here, as the minimum first touches the zone edge then proceeds to AFM3, but it remains theoretically possible if we carefully selected the parameters. Such cases will be vanishingly small regions of parameter space and hence we suggest the black regions are unlikely to be relevant for the practical study of FeTe.

\clearpage
\section{Reduction to a $4\times4$ Matrix \label{app:matReduction}}

After application of a Holstein-Primakoff transformation about the AFM3 ground state and Fourier transforming operators, we obtain a Hamiltonian  $H=\frac{S}{4}\sum_{\mathbf p} \mbox{\boldmath{$\psi$}}^\dagger \mathbf M \mbox{\boldmath{$\psi$}}$ up to an irrelevant constant term. $\mbox{\boldmath{$\psi$}}$ is the vector of creation and annihilation operators of the system:

\begin{equation}  
\mbox{\boldmath{$\psi$}}=
\begin{pmatrix*}[l]
\mbox{\boldmath{$\alpha$}}_{\mathbf p}^T &
\mbox{\boldmath{$\alpha$}}_{- \mathbf p}^T &
{\mbox{\boldmath{$\alpha$}}^{\dagger}}_{\mathbf p} &
{\mbox{\boldmath{$\alpha$}}^\dagger}_{-\mathbf p}
\end{pmatrix*}^T,
\end{equation}

\begin{equation}
\mbox{\boldmath{$\alpha$}}_{\mathbf p}=
\begin{pmatrix}
a_{\mathbf p} &
b_{\mathbf p} &
c_{\mathbf p} &
d_{\mathbf p}
\end{pmatrix}^T.
\end{equation}

The unit cell of the AFM3 phase has four sites, each of which has a creation and annihilation operator at $\mathbf p$ and $-\mathbf p$. As such, $\mathbf M$ is a $16\times 16$ matrix and $\mbox{\boldmath{$\psi$}}$ a vector of length 16; however, symmetries of the crystal enable an effective reduction in dimension.

The relevant symmetries are precisely the same as those present in the square (anti)-ferromagnet. In particular, there is always a reduction by at least a factor of four from what might be naively expected. One twofold reduction is due to the duplication of positive and negative eigenvalues. This is an unphysical distinction as each pair of positive and negative eigenvalues correspond to the same physical evolution as can be verified from analyzing a purely classical spin wave model --- i.e.\ magnetic moments imposing a torque on their neighbors. A further twofold reduction is due to the similarity of positive and negative momentum states. In the case of a ferromagnet this is most clear as there is no need to invoke negative momentum operators in the first place. This degeneracy remains even for an antiferromagnet, as it is dependent only on the inversion symmetry of the system. While a particular lattice may not be inversion symmetric, the underlying Bravais lattice always is and such a symmetry can be restored by a local phase transformation of the site operators.

The final twofold reduction is due to a physical symmetry, i.e.\ one that can actually be broken. This symmetry is composed of time reversal and a translation or rotation. The resulting degeneracy can be lifted in several equivalent ways. One may allow the spin magnitudes to vary between up and down sites (effectively breaking time reversal by giving the ground state a magnetization even for an AFM), introduce an external magnetic field, or break $U(1)$ symmetry in spin space by allowing $J_x$, $J_y$, and $J_z$ to differ. Any of these changes to the spin-wave Hamiltonian will lift one of the degeneracies in the simple cubic AFM or FeTe AFM3 cases by removing the ability to map some sites onto others in the unit cell.

Therefore, via a unitary transformation, it is possible to transform the $16\times 16$ matrix into a block diagonal matrix where each of the four $4\times4$ blocks is identical. For the dynamical matrix two of the matrices are the negative of the others but this is a minor detail and the eigenvalues are the same. This is the fourfold reduction which is always possible due to the above symmetries. It is therefore sufficient to calculate the eigenvalues of the $4\times4$ matrix knowing we expect to find two distinct branches. We will call this matrix $\tilde{\mathbf D}_4$ where the tilde indicates the basis has been transformed, $D$ rather than $M$ indicates we are using the dynamical matrix rather than coefficient matrix, and the subscript indicates the dimension. The relative intensity of the dispersion along each branch is calculated from the eigenvectors of the full dynamical matrix $\mathbf D_{16}$, which can be constructed from those of  $\tilde{\mathbf D}_4$ after ensuring orthonormality on a modified metric \cite{xiao2009theory}. In particular, the relative intensity is given by the magnitude of the sum of components of the eigenvector associated with the branch in question at each momentum. This makes intuitive sense if we view it as simply summing the contributions from each site in the unit cell to give the amplitude for a particular measurement.

Additionally, the unitary transformation which block diagonalizes the matrix is simply a reordering of the basis with phase adjustments. Therefore the magnitude of the sum of components of the length 16 vector is equal to the sum of components of the length 4 vectors so long as we get the phase information correct. We sketch out the process below for obtaining the intensity of a branch at a given momentum.

\begin{enumerate}
\item Find the eigenvalues and eigenvectors of $\tilde{\mathbf D}_4$.
\item Take the eigenvectors associated with the positive eigenvalues and normalise them on the metric 
\begin{equation}
\mathbf I_{4}^-=
\begin{pmatrix}
1 & 0 & 0 & 0\\
0 & 1 & 0 & 0\\
0 & 0 & -1 & 0\\
0 & 0 & 0 & -1
\end{pmatrix}.
\end{equation}
They will already be orthogonal on this metric.
\item Rephase the eigenvectors by left-multiplying by the matrix 
\begin{equation*}
\begin{pmatrix}
1 & 0 & 0 & 0\\
0 & e^{-i(p_y-q_y)a} & 0 & 0\\
0 & 0 &- e^{-2i(p_y-q_y)a} & 0\\
0 & 0 & 0 &- e^{-3i(p_y-q_y)a}
\end{pmatrix},
\end{equation*}
where $\mathbf p$ is the actual momentum under consideration and $\mathbf q$ is the center of the Brillouin zone that $\mathbf p$ is in. The $\mathbf p$ parts come from the rephasing in the unitary transformation (designed to reproduce the matrix used by Lipscombe \textit{et al}.\ \cite{lipscombe2011spin} for consistency) and the $\mathbf q$ terms come from adjusting for the Brillouin zone that the point is in. Together they make the relevant quantity how far from the center of its zone a point is in the $y$ direction. This is the relevant direction because in our chosen unit cell the sites are stacked along the $y$ axis.
\item Sum components of the vector and take the magnitude of the resulting complex number. 
\end{enumerate}

For completeness, we end by providing a collection of relations between the relevant matrices in this problem:
\begin{equation}
 \tilde{\mathbf D}_4=
\begin{pmatrix}
\epsilon_1 & J_{1b} f^* & - \epsilon_2 & -J_{1a} f \\
J_{1b} f & \epsilon_1 & -J_{1a} f^* & - \epsilon_2 \\
\epsilon_2 & J_{1a} & - \epsilon_1 & -J_{1a} f^* \\
J_{1a} f^* & \epsilon_2 & -J_{1b} f & -\epsilon_1
\end{pmatrix};
\end{equation}
\begin{equation}
\epsilon_1=2(J_{1a}+J_{2a}+2J_3-J_{1b}+J_{2b}(\cos(p_x a-p_y a)-1));
\end{equation}
\begin{equation}
\epsilon_2 = 2J_{2a} \cos(p_x a+p_y a)+2J_3(\cos(2 p_x a)+\cos(2 p_y a));
\end{equation}
\begin{equation}
f=e^{i p_x a}+e^{i p_y a};
\end{equation}
\begin{equation}
\tilde{\mathbf D}_{16}=
\begin{pmatrix}
\tilde{\mathbf D}_4& 0 & 0 & 0 \\
0 & \tilde{\mathbf D}_4 & 0 & 0 \\
0& 0 & -\tilde{\mathbf D}_4 & 0 \\
0 & 0 & 0 & -\tilde{\mathbf D}_4
\end{pmatrix}=\mathbf U^\dagger \mathbf D_{16}\mathbf U;
\end{equation}
\begin{equation}
\mathbf D_{16}=\mathbf I^-_{16} \mathbf M;
\end{equation}
\begin{equation}
\mathbf U = 
\begin{pmatrix}
\mbox{\boldmath{$\sigma$}}_x & 0 & 0 & 0 & 0 & 0 & 0 & 0 \\
0 & 0 & 0 & 0 & 0 & 0 & 0 & \mbox{\boldmath{$\sigma$}}_x\\
0 & 0 & 0 & 0 & 0 & \mbox{\boldmath{$\sigma$}}_x & 0 & 0 \\
0 & 0 &\mbox{\boldmath{$\sigma$}}_x & 0 & 0 & 0 & 0 & 0 \\ 
0 & 0 & 0  & \mbox{\boldmath{$\sigma$}}_x & 0 & 0 & 0 & 0 \\
0 & 0 & 0 & 0 & \mbox{\boldmath{$\sigma$}}_x & 0 & 0 & 0 \\
0 & 0 & 0 & 0 & 0 & 0 & \mbox{\boldmath{$\sigma$}}_x & 0 \\
0 & \mbox{\boldmath{$\sigma$}}_x & 0 & 0 & 0 & 0 & 0 & 0
\end{pmatrix}
\begin{pmatrix}
\mathbf R & 0 & 0 & 0\\
0 & \mathbf R & 0 & 0\\
0 & 0 &\mathbf R & 0\\
0 & 0 & 0 &\mathbf R
\end{pmatrix};
\end{equation}
\begin{equation}
\mathbf R=
\begin{pmatrix}
1 & 0 & 0 & 0\\
0 & e^{- i p_y a} & 0 & 0\\
0 & 0 &-e^{- i 2 p_y a} & 0\\
0 & 0 & 0 &-e^{- i 3 p_y a}
\end{pmatrix}.
\end{equation}
Here $\mathbf I^-_{16}=\mbox{\boldmath{$\sigma$}}_z \otimes \mathbf I_8$ is a diagonal $16\times 16$ matrix with $+1$ in the first eight diagonal entries and $-1$ in the final eight, and $\mbox{\boldmath{$\sigma$}}_x$ is the usual $2 \times 2$ Pauli-$x$ matrix.

\begin{acknowledgments}
CAH is grateful for financial support from UKRI via grant number EP/R031924/1.  This work was performed in part at Aspen Center for Physics, which is supported by National Science Foundation grant PHY-2210452.
\end{acknowledgments}

%

\end{document}